\documentclass{aa}
\usepackage[utf8]{inputenc}
\usepackage{graphicx}
\usepackage{txfonts}
\usepackage{verbatim}
\usepackage{siunitx}
\usepackage{hyperref}
\hypersetup{colorlinks, allcolors=blue}
\usepackage{subcaption}
\usepackage{comment}
\usepackage[labelfont=bf]{caption}
\usepackage{tablefootnote}
\usepackage[flushleft]{threeparttable}
\usepackage{caption}
\usepackage{amsmath}
\usepackage{relsize}
\usepackage{multirow}
\usepackage{mathtools}
\usepackage{orcidlink}

\usepackage{xcolor}
\newcommand{\gt}[1]{\textit{\textcolor{gray!70}{#1}}}

\usepackage[switch]{lineno}

\begin{document}

\title{The ESO SupJup Survey}
\subtitle{VIII. Chemical fingerprints of young L dwarf twins} 
\titlerunning{SupJup VIII: L dwarfs in AB Doradus}

\author{
N. Grasser\inst{\ref{instLeiden}}\orcidlink{0009-0009-6634-1741} \and 
I. A. G. Snellen\inst{\ref{instLeiden}}\orcidlink{0000-0003-1624-3667} \and 
S. de Regt\inst{\ref{instLeiden}}\orcidlink{0000-0003-4760-6168} \and 
D. Gonz\'alez Picos\inst{\ref{instLeiden}}\orcidlink{0000-0001-9282-9462} \and
Y. Zhang\inst{\ref{instCalTech}}\orcidlink{0000-0003-0097-4414} \and
T. Stolker\inst{\ref{instLeiden}}\orcidlink{0000-0002-5823-3072} \and
S. Gandhi\inst{\ref{instWarwick},\ref{instCEH}}\orcidlink{0000-0001-9552-3709} \and
E. Nasedkin\inst{\ref{instTCD}}\orcidlink{0000-0002-9792-3121} \and
R. Landman\inst{\ref{instLeiden}}\orcidlink{0000-0002-7261-8083} \and
A. Y. Kesseli\inst{\ref{instIPAC}}\orcidlink{0000-0002-3239-5989} \and
W. Mulder\inst{\ref{instLeiden}}\orcidlink{0000-0001-5762-9385}
}

\institute{
Leiden Observatory, Leiden University, Postbus 9513, 2300 RA, Leiden, The Netherlands \\
\email{grasser@strw.leidenuniv.nl} \label{instLeiden}
\and 
Department of Astronomy, California Institute of Technology, Pasadena, CA 91125, USA \label{instCalTech} \and
Department of Physics, University of Warwick, Coventry CV4 7AL, UK \label{instWarwick} \and
Centre for Exoplanets and Habitability, University of Warwick, Gibbet Hill Road, Coventry CV4 7AL, UK \label{instCEH} \and
School of Physics, Trinity College Dublin, University of Dublin, Dublin, Ireland \label{instTCD} \and
IPAC, Mail Code 100-22, Caltech, 1200 E. California Boulevard, Pasadena, CA 91125, USA \label{instIPAC}
}
\date{}

\abstract{The potentially distinct formation pathways of exoplanets and brown dwarfs may be imprinted in their elemental and isotopic ratios. This work is part of the ESO SupJup Survey, which aims to disentangle the formation pathways of super-Jupiters and brown dwarfs through the analysis of their chemical and isotopic ratios.} 
{In this study, we aim to characterize the atmospheres of two young L4 dwarfs, 2MASS J03552337+1133437 (2M0355) and 2MASS J14252798-3650229 (2M1425), in the AB Doradus Moving Group. This involved constraining their chemical composition, $^{12}$CO/$^{13}$CO ratio, pressure-temperature profile, surface gravity, and rotational velocity, among other parameters.} 
{We have obtained high-resolution CRIRES+ K-band spectra of these brown dwarfs, which we analyzed with an atmospheric retrieval pipeline. Atmospheric models were generated with the radiative transfer code \texttt{petitRADTRANS}, for which we employed a free and equilibrium chemistry approach, which we coupled with the \texttt{PyMultiNest} sampling algorithm to determine the best fit.}
{We report robust detections of $^{13}$CO (13.4 \& 8.0\,$\sigma$) and HF (11.6 \& 15.8\,$\sigma$) in 2M0355 and 2M1425, respectively. The objects have similar overall atmospheric properties, including $^{12}$CO/$^{13}$CO isotope ratios of $95.5\substack{+6.8 \\ -6.4}$ for 2M0355 and $109.6\substack{+10.6 \\ -9.6}$ for 2M1425. The most notable difference is the robust evidence of CH$_4$ (5.5\,$\sigma$) but no H$_2$S (<\,2.3\,$\sigma$) in 2M1425, in contrast to 2M0355, for which we retrieved H$_2$S (4.6\,$\sigma$) but not CH$_4$ (<\,2.2\,$\sigma$). We also find tentative hints of NH$_3$ in 2M1425 (3.0\,$\sigma$). In both brown dwarfs, we find only tentative hints of H$_2^{18}$O (1.1 \& 3.0\,$\sigma$), with lower limits of H$_2^{16}$O/H$_2^{18}$O\,$\sim$\,1000. Both objects appear to be close to chemical equilibrium, considering the main spectral contributors. We also find that 2M1425 is 50--200~K hotter than 2M0355 and has a higher surface gravity.} 
{Retrievals of high-resolution K-band spectra of young brown dwarfs enable a detailed insight into their atmospheres. As with similar targets, these brown dwarfs also exhibit a depletion of $^{13}$CO compared to the interstellar medium. Future studies will put them into the context of other objects observed as part of the ESO SupJup Survey.}


\maketitle

\section{Introduction}
Atmospheric spectra of exoplanets and brown dwarfs are key to constraining their composition, climate, and chemical and physical processes. Their formation and evolutionary pathways may be encoded within their chemical constituents and isotopic ratios, since the chemical composition of solids and gases are expected to depend on their birth environment \citep{Turrini2021, Pacetti2022}. Among the proposed tracers of formation are elemental ratios such as C/O (e.g., \citealt{Oberg2011}) and isotope ratios (e.g., \citealt{MolliereSnellen2019}). Objects that formed via gravitational collapse are expected to have a $^{12}$C/$^{13}$C ratio similar to the parent cloud from which they formed \citep{Oberg2021}. Such top-down formation is expected for isolated brown dwarfs, similar to star formation \citep{Bate2002}. On the other hand, objects that formed via core accretion are thought to inherit the $^{12}$C/$^{13}$C ratio of the local disk solids, which is expected to vary throughout the disk due to isotope fractionation processes \citep{Visser2009, Yoshida2022}. Although planets are typically presumed to form via core accretion, the postulation is less clear for larger companion objects, such as giant exoplanets and companion brown dwarfs, especially those on wide orbits \citep{Bergin2024}. Giant exoplanets can form via core accretion \citep{Pollack1996}, though they may also form via gravitational instability \citep{Boss1997, Mayer2002}. In this regard, atmospheric tracers are essential to establishing a better understanding of these formation processes.

Considering their higher luminosity, observations of brown dwarfs offer a more accessible method for investigating the chemical and physical processes that govern the atmospheres of hot giant planets, due to similarities in chemical composition, temperature, and surface gravity \citep{Patience2012, Faherty2016}. The advancement of high-resolution spectroscopy (HRS) and analysis techniques have enabled unprecedented insights into their atmospheric composition, thermal structure, potential clouds, and other characteristics (e.g., \citealt{Whiteford2023, Landman2024}). The first $^{12}$C/$^{13}$C measurement in a substellar companion (YSES 1b) yielded a ratio of $^{12}$CO/$^{13}$CO\,=\,$31\substack{+17 \\ -10}$ \citep{Zhang2021_SJ} with medium resolution spectroscopy. An enrichment in $^{13}$C could stem from the accretion of $^{13}$C-rich ices beyond the CO-snowline. However, new data with HRS suggests a ratio of $^{12}$CO/$^{13}$CO\,=\,$88\pm13$ \citep{Zhang2023}, which is consistent with the primary's isotope ratio. The measured $^{12}$C/$^{13}$C ratios of brown dwarfs \citep{Zhang2021_BD, Costes2024, deRegt2024} have so far been above or around that of the interstellar medium (ISM), which has a ratio of $^{12}$C/$^{13}$C\,$=68 \pm 15$ \citep{Milam2005}. Emerging trends supported by a larger sample of objects may show distinct mechanisms governing the formation and evolution of brown dwarfs and giant exoplanets, resulting in different atmospheric isotopolog ratios.

\begin{figure}[t!]
    \centering
    \includegraphics[width=\linewidth]{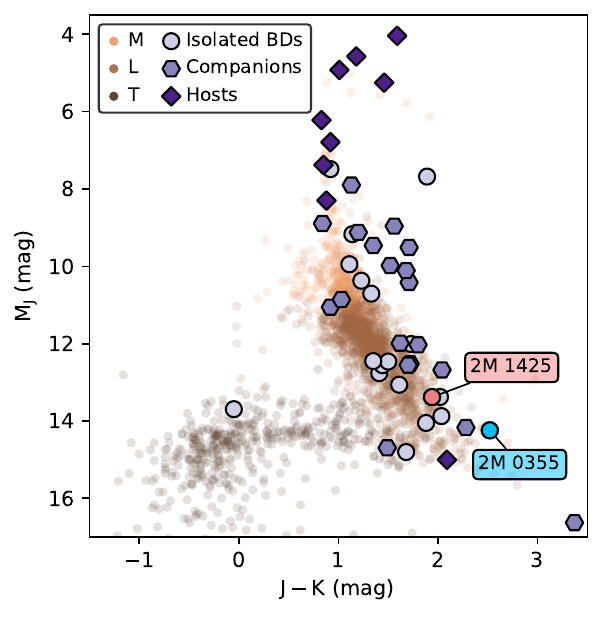}
    \caption{Color-magnitude diagram showing the sample of objects observed throughout the SupJup Survey in shades of purple, categorized into isolated brown dwarfs (BDs), companions, and hosts. The objects of this study, 2M0355 and 2M1425, are highlighted in blue and pink, respectively. Photometry of isolated brown dwarfs from the UltracoolSheet (\url{https://doi.org/10.5281/zenodo.13993077}) are shown in the background for reference, with their colors representing their spectral types.}
    \label{fig:CMD}
\end{figure}

\begin{table}[t!]
    \centering
    \caption{Properties and our observing setup of the brown dwarfs 2M0355 and 2M1425.}
    \resizebox{\linewidth}{!}{
    \begin{tabular}{lcc}
    \hline
    \hline
     Parameter & 2M0355 & 2M1425  \\
     \hline
     Spectral type (optical/IR) &  L5$\gamma$$^{(\text{a})}$/L4$\gamma$$^{(\text{b})}$ & L3$^{(\text{c})}$/L4$\gamma$$^{(\text{d})}$ \\
     Radius [$R_{\text{Jup}}$] & 1.20$\,\pm\, 0.04^{(\text{d})}$ & 1.23$\,\pm\, 0.01^{(\text{d})}$ \\
     Mass [$M_{\text{Jup}}$] & $23\substack{+2\\-5}^{(\text{e})}$ & $22\substack{+3\\-8}^{(\text{e})}$ \\
     Effective temperature [K] & 1478$\,\pm\,57^{(\text{f})}$
     & 1535$\,\pm\,55^{(\text{f})}$
     \\
     Surface gravity [cgs] & 4.67\,$\pm\, 0.05^{(\text{d})}$ & 4.64\,$\pm\,0.04^{(\text{d})}$ \\
     Radial velocity [km/s] & 11.92$\,\pm\,0.22^{(\text{d})}$ & 5.37$\,\pm\,0.25^{(\text{g})}$ \\
     Rotational velocity [km/s] & 
     <\,4$^{(\text{h})}$ & 32.37$\,\pm\,0.66^{(\text{g})}$ \\
     Age [Myr] & 125$^{(\text{e})}$ & 125$^{(\text{e})}$ \\
     Distance [pc] & 9.16$\,\pm\,0.04^{(\text{i})}$ &  11.85$\,\pm\,0.04^{(\text{i})}$\\
     $K_{\text{S}}$ [mag] & 11.53$\,\pm\,0.02^{(\text{j})}$ & 11.81$\,\pm\,0.03^{(\text{j})}$ \\
     \hline
     Observation date & 2022-11-02 & 2023-02-01 \\
     Slit width ["] & 0.2 & 0.4 \\
     Exposure time [min] & 50  & 60 \\
     Seeing ["] & 0.73\,$\pm$\,0.03 & 0.72\,$\pm$\,0.03 \\
     Airmass & 1.42\,$\pm$\,0.05 & 1.13\,$\pm$\,0.04 \\
     Integrated water vapor [mm] & 1.29\,$\pm$\,0.01 & 13.66\,$\pm$\,0.04 \\
     Signal-to-noise ratio & 30 & 40 \\
     \hline
    \end{tabular}
    }
    \tablefoot{References:  $^{(\text{a})}$\cite{Cruz2009}, $^{(\text{b})}$\cite{Bardalez2019},$^{(\text{c})}$\cite{Reid2008}, $^{(\text{d})}$\cite{Gagne2015}, $^{(\text{e})}$\cite{Aller2016}, $^{(\text{f})}$\cite{Filippazzo2015}, $^{(\text{g})}$\cite{Blake2010}, $^{(\text{h})}$\cite{Zhang2021_BD}, $^{(\text{i})}$\cite{Gaia2020}, $^{(\text{j})}$\cite{2MASS2003}}
    \label{tab:properties}
\end{table}

In this regard, the ESO SupJup Survey (Program ID: 1110.C-4264, PI: Snellen) aims to disentangle the formation pathways of super-Jupiters, substellar companions, and isolated brown dwarfs through their chemical composition and isotopic ratios \citep{deRegt2024}. High-resolution K- and J-band spectra of 19 isolated objects, 19 lower-mass companions, and 11 hosts have been obtained with the upgraded CRyogenic high-resolution InfraRed Echelle Spectrograph (CRIRES+) on the Very Large Telescope (VLT) at the Paranal Observatory in Chile (e.g., \citealt{Dorn2014, Dorn2023}). The first results of several brown dwarfs yielded $^{12}$CO/$^{13}$CO ratios above or around that of the local ISM \citep{deRegt2024, Picos2024, Mulder2025}, while results for substellar companions exhibited more variation from enrichment to depletion in $^{13}$CO \citep{Zhang2024, Picos2025, Gandhi2025}. This work is part of the ongoing analysis of the data obtained throughout the survey, and our aim is to characterize the two isolated brown dwarfs 2MASS J03552337+1133437 (2M0355) and 2MASS J14252798-3650229 (2M1425), which both belong to the AB Doradus Moving Group \citep{Gagne2018}. Figure~\ref{fig:CMD} shows the color-magnitude diagram of the diverse sample of SupJup objects, which range from isolated brown dwarfs to companions and hosts, highlighting the positions of 2M0355 and 2M1425.

The L dwarf 2M0355 was discovered by \cite{Reid2006} and is among the nearest and reddest brown dwarfs observed \citep{Reid2008}. Compared to the spectral standard, its spectrum exhibits peculiar features that are indicative of a low surface gravity, such as a reduced absorption of the KI doublet and other alkali metals. This is denoted in its classification as a L5$\gamma$ dwarf \citep{Cruz2009}, where $\gamma$ stands for low gravity. Since young brown dwarfs are still contracting and consequently have larger radii and a lower surface gravity, this implies that 2M0355 is young ($\sim$\,100\,Myr, \citealt{Cruz2009}). Using near infrared (NIR) spectroscopy, \cite{Faherty2013} found that 2M0355 is indeed redder and less luminous compared to typical brown dwarfs, similar to directly imaged massive exoplanets, which could be explained by photospheric dust in its atmosphere. The similar shape of its H-band spectrum compared to that of a 10~Myr L dwarf and its membership of the AB Doradus moving group further supports that 2M0355 is a young object ($\sim$\,125\,Myr, \citealt{Liu2013, Aller2016}). Using high-resolution K-band Keck/NIRSPEC data of 2M0355, \cite{Zhang2021_BD} retrieved an isotopolog ratio of $^{12}$CO/$^{13}$CO\,=\,$97\substack{+25 \\ -18} $, which is higher than the interstellar standard ($\sim$\,68). Furthermore, CRIRES+ K-band observations of 2M0355, carried out by \cite{Zhang2022} as part of the Science Verification Observations (with a total exposure time of 20\,min compared to 50\,min in this study) yielded a carbon isotope ratio of $^{12}$C/$^{13}$C\,=\,$108 \pm 10$ and a tentative measurement of the oxygen isotope ratio of $^{16}$O/$^{18}$O\,=\,$ 1489 \substack{+1027\\ -426}$.

The second brown dwarf we investigate in this study, 2M1425, was discovered by \cite{Kendall2004} during a search for nearby ultracool dwarfs in the Deep Near-Infrared Survey of the Southern Sky (DENIS) catalog. It is classified as a L4$\gamma$ dwarf. Found to be a member of AB Doradus \citep{Gagne2015_BANYAN}, it is also a young object and estimated to have a similar age as 2M0355. Table~\ref{tab:properties} summarizes the main characteristics of 2M1425 and 2M0355. Both objects have very similar physical properties, with a radius of $R$\,$\sim$\,$1.2\,R_{\text{Jup}}$, mass of $M$\,$\gtrapprox$\,20\,$M_{\text{Jup}}$, effective temperature of $T_\text{eff}$\,$\sim$\,1500\,K, and surface gravity of log\,$g$\,$\sim$\,4.6 \citep{Gagne2015, Aller2016}. Despite their bulk resemblances, \cite{Gagne2015} report that their NIR spectral shapes exhibit notable differences, among which is the significantly redder continuum of 2M0355 as well as its weaker FeH and CO absorption. This disparity could arise from different cloud properties or the presence of an unresolved companion.

In this study, we characterize the atmospheres of the isolated brown dwarfs 2M0355 and 2M1425 through a Bayesian atmospheric retrieval framework. This method infers probability distributions of atmospheric properties, allowing us to obtain an atmospheric model that best explains the observations. In Section~\ref{sec:obs}, we summarize the observational setup and data reduction. Section~\ref{sec:methods} provides an overview of the retrieval framework, which describes the properties of the model atmospheric spectra and the likelihood evaluation. In Section~\ref{sec:results}, we present and discuss our results, concluding our work in Section~\ref{sec:concl}.

\section{Observations} \label{sec:obs}

\begin{figure*}[ht!]
    \centering
    \includegraphics[width=\linewidth]{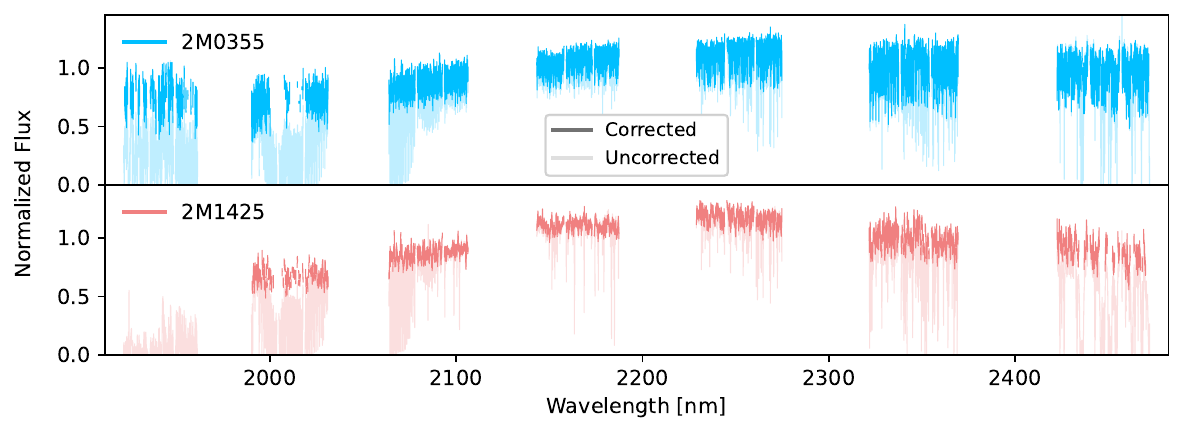}
    \caption{Observed spectra of 2M0355 (top panel) and 2M1425 (bottom panel). The transparent lines show the reduced observations before correcting for tellurics with \texttt{Molecfit}, while the opaque lines show the telluric-corrected spectra that are used in our retrievals. The bluest order of 2M1425 was masked entirely due to strong atmospheric absorption.}
    \label{fig:observations}
\end{figure*}

The two brown dwarfs, 2M0355 and 2M1425, were observed with VLT/CRIRES+ in the K-band as part of the ESO SupJup Survey on the night of November 2, 2022, and February 1, 2023, respectively. The observations were conducted using ABBA nodding sequences, with a nod throw of 6" and 5", resulting in 10 and 12 exposures of 300\,s each for 2M0355 and 2M1425. No adaptive optics were used. Both objects were observed in the K2166 wavelength setting, covering the wavelength range from 1.92 to 2.48\,$\mu$m, including the $^{13}$CO bandhead near 2.345~$\mu$m. The two objects were observed using different slit widths based on their projected rotational velocities ($v$\,sin\,$i$) known from the literature (see Table~\ref{tab:properties}). As 2M0355 has a $v$\,sin\,$i$\,<\,4 \citep{Zhang2021_BD}, and hence a low rotational broadening, we opted for the narrow-slit mode of 0.2" to acquire a higher resolution. On the other hand, 2M1425, with a $v$\,sin\,$i$\,>\,30 \citep{Blake2010}, was observed with the wide-slit mode of 0.4" to obtain a higher signal-to-noise ratio (S/N). Due to its intrinsically large rotational broadening, the observations of 2M1425 are not limited by instrumental broadening.

The telluric standard stars $\lambda$ Tau and $\beta$ Hya were observed in the same settings for 2M0355 and 2M1425, respectively. Observing conditions were stable, with an average seeing of 0.7 for both objects, and the airmass did not vary greatly during the observations. In contrast to 2M0355, the humidity during the observations of 2M1425 was very high, with an average integrated water vapor of 13.7\,mm. Relevant parameters of the observational setup and conditions are listed in the lower part of Table~\ref{tab:properties}. The data reduction was carried out with \texttt{excalibuhr}\footnote{\url{https://github.com/yapenzhang/excalibuhr}} \citep{Zhang2024}, a Python package for data reduction of CRIRES+ spectroscopy based on methods used in \cite{HolmbergMadu2022} and \texttt{pycrires} \citep{StolkerLandman2023}. The reduction pipeline includes flat-fielding, spectral order tracing, blaze correction, sky background subtraction, nodding pair combination, spectral extraction, and wavelength solution calibration. More details can be found in \cite{deRegt2024}, who use the same data reduction routine as we did.  

Observations obtained using ground-based telescopes are subject to contamination by the Earth’s atmospheric transmission. To correct for these tellurics, the software \texttt{Molecfit} \citep{Smette2015} was applied to the standard star spectra. \texttt{Molecfit} fits a synthetic model of the Earth’s atmospheric transmission to the observations, which is generated based on the atmospheric conditions during the observations. This includes the airmass, precipitable water vapor, thermal profile and main opacity sources, such as H$_2$O, CO$_2$, CO, CH$_4$, and N$_2$O. For this process, we use our observations of the telluric standard stars $\lambda$ Tau and $\beta$ Hya for 2M0355 and 2M1425, respectively, which provide a clean measurement of the tellurics and instrumental throughput. The observed spectrum is divided by the telluric model obtained with \texttt{Molecfit}, which results in the telluric-corrected spectrum that we use in our retrievals. Pixels containing the deepest tellurics, with a flux below 70\% of the normalized continuum, are masked. The instrumental throughput is recovered by dividing the blackbody spectrum of the respective standard star, namely $T_{\text{eff}}$\,=\,15\,536\,K for $\lambda$ Tau \citep{Royer2024} and $T_{\text{eff}}$\,=\,10\,980\,K for $\beta$ Hya \citep{Arcos2018}.

Figure~\ref{fig:observations} shows the original reduced observations and telluric-corrected spectra for 2M0355 and 2M1425. The spectra consist of seven spectral orders, whose wavelength regions are spread over three detectors with 2048 pixels each, which creates gaps in the observed wavelength range. To simplify subsequent steps, the spectra are normalized by the median flux. The high humidity during the observations of 2M1425 caused significantly stronger telluric absorption, particularly at shorter wavelengths. Consequently, the entire bluest order of 2M1425 had to be excluded from our analysis. The spectra have a signal-to-noise ratio of approximately 30 (2M0355) and 40 (2M1425) per pixel at 2.345\,$\mu$m\footnote{To determine the S/N, the flux uncertainties are multiplied with the uncertainty scaling factor $s^2$ from Equation~\ref{equ:s2} from Section~\ref{subsec:lncov}, calculated from the best-fit model. This provides a more reliable estimate of the noise.}.

As the line profiles of the standard star were fit with a Gaussian, the resolving power $\mathcal{R}$ is determined through the full width at half maximum. According to the telluric line fits, our data exhibits a spectral resolution of $\mathcal{R}$\,$\sim$\,$90.000$ for 2M0355 and $\mathcal{R}$\,$\sim$\,$60.000$ for 2M1425, which align with the nominal resolution of CRIRES+ in the narrow and wide slit mode, respectively. Through the positions of the telluric absorption lines, \texttt{Molecfit} also calibrates the wavelength solution of the observations. However, comparing a preliminary model atmosphere (including mainly H$_2$O and CO) with the spectrum of 2M0355, we noticed that its wavelength solution exhibited slight inaccuracies in the reddest order-detector pair. We refined the wavelength solution using a quadratic polynomial correction, achieved by maximizing cross-correlation functions between the data and model spectrum \citep{Zhang2024}.

\section{Retrieval framework} \label{sec:methods}

\begin{table*}[]
    \centering
    \caption{Free parameters, prior ranges, and retrieval results for 2M0355 and 2M1425.}
    \resizebox{\textwidth}{!}{
    \begin{tabular}{lll|rrrr}
    \hline
    \hline
    Parameter & Description & Prior range & \multicolumn{2}{c}{2M0355} & \multicolumn{2}{c}{2M1425} \\
    &&& Free & Equ & Free & Equ \\
    \hline
$v_{\text{rad}}$ [km/s] & Radial velocity & $\mathcal{U}$(2,20) & 13.25$^{+0.01}_{-0.01}$  & 13.25$^{+0.01}_{-0.01}$  & 5.54$^{+0.05}_{-0.05}$  & 5.54$^{+0.06}_{-0.06}$ \\
$v\,\text{sin}\,i$ [km/s] & Projected rotational velocity & $\mathcal{U}$(0,40) & 3.04$^{+0.08}_{-0.08}$  & 3.06$^{+0.10}_{-0.10}$  & 31.57$^{+0.19}_{-0.23}$  & 31.73$^{+0.31}_{-0.29}$ \\
log $g$ [cm/s$^2$] & Surface gravity & $\mathcal{U}$(3,5) & 4.75$^{+0.03}_{-0.03}$  & 4.73$^{+0.04}_{-0.04}$  & 4.98$^{+0.01}_{-0.01}$  & 4.88$^{+0.03}_{-0.02}$ \\
$\epsilon_\mathrm{limb}$ & Limb-darkening coefficient & $\mathcal{U}$(0.2,1) & 0.72$^{+0.14}_{-0.16}$  & 0.73$^{+0.16}_{-0.21}$  & 0.74$^{+0.04}_{-0.05}$  & 0.76$^{+0.06}_{-0.06}$ \\
\hline
log H$_2$O & log$_{10}$ VMR of H$_2$O & $\mathcal{U}$(-12,-1) &  -3.00$^{+0.03}_{-0.03}$  & \gt{-3.02$^{+0.06}_{-0.04}$}  & -3.20$^{+0.01}_{-0.01}$  & \gt{-3.26$^{+0.01}_{-0.01}$} \\
log $^{12}$CO & log$_{10}$ VMR of $^{12}$CO & $\mathcal{U}$(-12,-1) & -2.74$^{+0.03}_{-0.03}$  & \gt{-2.75$^{+0.04}_{-0.05}$}  & -2.92$^{+0.01}_{-0.01}$  & \gt{-3.05$^{+0.03}_{-0.01}$} \\
log $^{13}$CO & log$_{10}$ VMR of $^{13}$CO & $\mathcal{U}$(-12,-1) &  -4.72$^{+0.04}_{-0.04}$  & \gt{-4.74$^{+0.08}_{-0.07}$}  & -4.96$^{+0.04}_{-0.04}$  & \gt{-5.03$^{+0.09}_{-0.09}$} \\
log C$^{18}$O & log$_{10}$ VMR of C$^{18}$O & $\mathcal{U}$(-12,-1) & -7.96$^{+1.60}_{-1.91}$  & \gt{-10.69$^{+3.47}_{-2.50}$}  & -9.22$^{+1.38}_{-1.31}$  & \gt{-10.94$^{+1.59}_{-1.71}$} \\
log C$^{17}$O & log$_{10}$ VMR of C$^{17}$O & $\mathcal{U}$(-12,-1) & -9.05$^{+1.39}_{-1.52}$  & \gt{-9.93$^{+2.25}_{-2.63}$}  & -8.44$^{+1.18}_{-1.40}$  & \gt{-9.27$^{+1.67}_{-3.14}$} \\
log CH$_4$ & log$_{10}$ VMR of CH$_4$ & $\mathcal{U}$(-12,-1) & -9.30$^{+1.14}_{-1.19}$  & \gt{-5.95$^{+0.05}_{-0.03}$}  & -5.35$^{+0.05}_{-0.08}$  & \gt{-6.50$^{+0.03}_{-0.02}$} \\
log NH$_3$ & log$_{10}$ VMR of NH$_3$ & $\mathcal{U}$(-12,-1) &  -8.98$^{+1.13}_{-1.26}$  & \gt{-6.34$^{+0.02}_{-0.02}$}  & -6.49$^{+0.25}_{-0.56}$  & \gt{-6.46$^{+0.01}_{-0.01}$} \\
log HCN & log$_{10}$ VMR of HCN & $\mathcal{U}$(-12,-1) & -8.70$^{+1.76}_{-1.54}$  & \gt{-7.52$^{+0.02}_{-0.02}$}  & -8.22$^{+1.20}_{-1.19}$  & \gt{-7.42$^{+0.01}_{-0.02}$} \\
log HF & log$_{10}$ VMR of HF & $\mathcal{U}$(-12,-1) &  -6.96$^{+0.06}_{-0.06}$  & \gt{-6.77$^{+0.04}_{-0.04}$}  & -7.09$^{+0.04}_{-0.04}$  & \gt{-7.13$^{+0.04}_{-0.01}$} \\
log H$_2$S & log$_{10}$ VMR of H$_2$S & $\mathcal{U}$(-12,-1) &  -4.15$^{+0.06}_{-0.07}$  & \gt{-4.06$^{+0.04}_{-0.04}$}  & -9.00$^{+1.30}_{-1.27}$  & \gt{-4.48$^{+0.03}_{-0.01}$} \\
log H$_2^{18}$O & log$_{10}$ VMR of H$_2^{18}$O & $\mathcal{U}$(-12,-1) & -7.34$^{+1.51}_{-3.06}$  & \gt{-10.94$^{+3.44}_{-2.49}$}  & -5.90$^{+0.20}_{-3.06}$  & \gt{-11.15$^{+1.60}_{-1.69}$} \\
\hline
C/O & Carbon-to-oxygen ratio & $\mathcal{U}$(0,1) &  \gt{0.65$^{+0.01}_{-0.01}$}  & 0.52$^{+0.00}_{-0.00}$  & \gt{0.66$^{+0.00}_{-0.00}$}  & 0.57$^{+0.00}_{-0.01}$ \\
$[$Fe/H$]$ \gt{or [C/H]} & Metallicity & $\mathcal{U}$(-1.5,-1.5) &  \gt{0.58$^{+0.03}_{-0.03}$}  & 0.60$^{+0.04}_{-0.04}$  & \gt{0.40$^{+0.01}_{-0.01}$}  & 0.26$^{+0.01}_{-0.01}$ \\
log $^{12}$C/$^{13}$C & log$_{10}$ ratio of $^{12}$C to $^{13}$C & $\mathcal{U}$(1,12) & \gt{1.98$^{+0.03}_{-0.03}$}  & 1.97$^{+0.04}_{-0.04}$  & \gt{2.04$^{+0.04}_{-0.04}$}  & 1.99$^{+0.05}_{-0.05}$ \\
log $^{16}$O/$^{17}$O & log$_{10}$ ratio of $^{16}$O to $^{17}$O & $\mathcal{U}$(1,12) & \gt{6.31$^{+1.53}_{-1.40}$}  & 8.03$^{+2.19}_{-2.21}$  & \gt{5.52$^{+1.39}_{-1.17}$}  & 6.50$^{+2.86}_{-2.57}$ \\
log $^{16}$O/$^{18}$O & log$_{10}$ ratio of $^{16}$O to $^{18}$O & $\mathcal{U}$(1,12) & \gt{4.34$^{+3.09}_{-1.51}$}  & 3.20$^{+3.30}_{-0.20}$  & \gt{2.71$^{+3.05}_{-0.20}$}  & 5.99$^{+3.01}_{-2.59}$ \\
\hline
$T_0$ [K] & Temperature at $P_0=10^2\,$bar& $\mathcal{U}$(1000,4000) & 2399.23$^{+113.53}_{-93.91}$  & 2701.89$^{+150.57}_{-144.17}$  & 3833.87$^{+62.89}_{-59.62}$  & 3963.03$^{+24.85}_{-41.09}$ \\
 $\nabla T_0$ & Gradient at $P_0=10^2\,$bar& $\mathcal{U}$(0,0.4) & 0.19$^{+0.09}_{-0.10}$  & 0.02$^{+0.03}_{-0.02}$  & 0.28$^{+0.06}_{-0.07}$  & 0.17$^{+0.11}_{-0.09}$ \\
$\nabla T_1$ & Gradient at $P_1=10^0\,$bar& $\mathcal{U}$(0,0.4) & 0.24$^{+0.02}_{-0.02}$  & 0.15$^{+0.01}_{-0.01}$  & 0.26$^{+0.05}_{-0.05}$  & 0.37$^{+0.02}_{-0.02}$ \\
$\nabla T_2$ & Gradient at $P_2=10^{-2}\,$bar& $\mathcal{U}$(0,0.4) & 0.04$^{+0.00}_{-0.00}$  & 0.04$^{+0.00}_{-0.00}$  & 0.05$^{+0.01}_{-0.01}$  & 0.05$^{+0.00}_{-0.00}$ \\
$\nabla T_3$ & Gradient at $P_3=10^{-4}\,$bar& $\mathcal{U}$(0,0.4) & 0.05$^{+0.00}_{-0.00}$  & 0.05$^{+0.00}_{-0.00}$  & 0.08$^{+0.00}_{-0.00}$  & 0.09$^{+0.00}_{-0.00}$ \\
$\nabla T_4$ & Gradient at $P_4=10^{-6}\,$bar& $\mathcal{U}$(0,0.4) & 0.15$^{+0.03}_{-0.02}$  & 0.22$^{+0.03}_{-0.03}$  & 0.29$^{+0.01}_{-0.01}$  & 0.28$^{+0.01}_{-0.01}$ \\
\hline
log $\kappa_{\mathrm{cl},0}$ [cm$^2$/g] & Opacity at cloud base & $\mathcal{U}$(-10,3) &  -5.78$^{+2.24}_{-2.12}$  & -4.88$^{+2.90}_{-2.62}$  & -4.17$^{+1.83}_{-1.82}$  & -5.27$^{+2.66}_{-2.55}$ \\
log $P_{\mathrm{cl},0}$ [bar] & Cloud base pressure & $\mathcal{U}$(-6,3) & -2.27$^{+2.03}_{-1.88}$  & -2.09$^{+2.64}_{-2.09}$  & -1.45$^{+2.52}_{-2.32}$  & -1.82$^{+2.44}_{-2.20}$ \\
$f_\mathrm{sed}$ & Cloud decay power & $\mathcal{U}$(0,20) & 11.35$^{+3.92}_{-4.14}$  & 10.54$^{+4.79}_{-5.26}$  & 8.30$^{+4.60}_{-4.08}$  & 11.48$^{+4.67}_{-5.56}$ \\
log $a$ & GP amplitude & $\mathcal{U}$(-1,1) & 0.28$^{+0.00}_{-0.00}$  & 0.28$^{+0.00}_{-0.00}$  & 0.14$^{+0.00}_{-0.00}$  & 0.14$^{+0.00}_{-0.00}$ \\
log $l$ [nm] & GP length-scale & $\mathcal{U}$(-3,0) & -1.76$^{+0.00}_{-0.00}$  & -1.76$^{+0.00}_{-0.00}$  & -1.91$^{+0.01}_{-0.01}$  & -1.91$^{+0.01}_{-0.01}$ \\
\hline
$\chi^2$ & $\chi^2$ of the best-fit model & & 49.78  & 49.85  & 17.43  & 17.53 \\
ln$\,\mathcal{L}_{\text{free}}\,$--\,ln$\,\mathcal{L}_{\text{equ}}$ & \multicolumn{2}{l|}{Log-likelihood (Equ.~\ref{equ:lnL}) difference (free vs. equ)} & \multicolumn{2}{c}{11.88} & \multicolumn{2}{c}{33.01} \\
ln\,$B_m$ ($\sigma$) & \multicolumn{2}{l|}{Bayes factor (significance) betw. free vs. equ} & \multicolumn{2}{c}{-6.66 (-4.07$\,\sigma$)} & \multicolumn{2}{c}{-2.67 (-2.82$\,\sigma$)} \\
\hline
    \end{tabular}
    }
    \tablefoot{The three columns on the left show the parameter notation, description and prior ranges. $\mathcal{U}$ denotes a uniform distribution within the given range. The four rightmost columns show the retrieval results for 2M0355 and 2M1425, including their 1\,$\sigma$ errors, using the free and equilibrium (equ) chemistry setup. The gray cursive values were not retrieved as free parameters, but determined from the results. The abundances of the equilibrium chemistry retrievals are given at the pressure where the integrated emission contribution is maximal. The shown $^{16}$O/$^{18}$O ratio for the free chemistry setup is based on H$_2$O. The logarithm of the Bayes factor ln\,$B_m$ and significance $\sigma$ are determined from the ln\,$\mathcal{Z}$ (nested importance sampling global log-evidence, outputted by \texttt{PyMultiNest}) with respect to the free chemistry retrieval.}
    \label{tab:free_params}
\end{table*}

Atmospheric retrievals typically consist of two frameworks, namely one to forward model a spectrum based on certain physical parameters, and one to iteratively sample the prior space of each parameter to obtain a best-fitting model that matches the observed data. Our atmospheric retrieval utilizes the nested sampling tool \texttt{PyMultiNest} \citep{Buchner2014}, which is a Python wrapper of the Bayesian inference algorithm \texttt{MultiNest} \citep{Feroz2009}, to sample the parameter space. Model spectra were generated with the radiative transfer code \texttt{petitRADTRANS} (pRT; Version 2.7; \citealt{Molliere2019}) based on various parameters that influence the atmospheric spectrum, such as the chemical composition, thermal profile, cloud properties, and surface gravity, which are retrieved as free parameters. Table~\ref{tab:free_params} lists all of the free parameters we include in our retrieval, including their descriptions and prior ranges. To avoid biasing the results, we opted for very wide priors to sample a large part of the parameter space. The retrievals were run in parallel mode on the Leiden Exoplanet Machine (\texttt{LEM}) using 40 CPUs. We followed the \texttt{MultiNest} recommendations and use the Importance Nested Sampling (INS) mode with a sampling efficiency of 5\% \citep{Feroz2019}. Furthermore, we applied the constant efficiency mode for faster convergence. For both objects, the posterior distribution was sampled with 400 live points, and the retrieval was run until an evidence tolerance of 0.5 was reached, as in \cite{deRegt2024}.

\subsection{Model spectra}

The emission spectra used as models in our retrieval are generated with \texttt{petitRADTRANS} according to the atmospheric parameters. To allow for faster computations, the opacities are down-sampled by a factor of three from the intrinsic resolution of $\mathcal{R}$\,=\,$10^6$. The model spectra are shifted to the barycentric rest frame using the \texttt{helcorr} function from Astropy \citep{Astropy2022}. Afterwards, they are broadened according to the projected rotational velocity $v\,\text{sin}\,i$ and the limb-darkening coefficient $\epsilon_\text{limb}$ using the \texttt{fastRotBroad} function from \texttt{PyAstronomy}\footnote{\url{https://github.com/sczesla/PyAstronomy}} \citep{Gray2008, Czesla2019}, all three of which are retrieved as free parameters. The model spectra are then down-convolved to match the resolution of the data.

To avoid making any assumptions on the cloud composition, we implement a simple gray cloud model, as in \cite{deRegt2024}. The cloud opacity $\kappa_\mathrm{cl}(P)$ at a pressure $P$ is calculated through the opacity at the cloud base $\kappa_{\mathrm{cl},0}$, with the cloud base being located at a atmospheric certain pressure $P_{\mathrm{cl},0}$. The decay of the cloud opacity above the base is defined through a power-law using the sedimentation parameter $f_\text{sed}$, while the cloud opacity is set to zero below the base. Our retrieval fits for the parameters $\kappa_{\mathrm{cl},0}$, $P_{\mathrm{cl},0}$, and $f_\text{sed}$, with which the cloud opacity is expressed as

\begin{equation}
    \kappa_\mathrm{cl}(P) = 
    \begin{cases}
      \kappa_{\mathrm{cl},0} \left(\dfrac{P}{P_\mathrm{base}}\right)^{f_\mathrm{sed}} & P<P_\mathrm{base} \\
      0 & P\geq P_\mathrm{base} \\
    \end{cases}
    .
\end{equation}

\subsubsection{Chemistry}

\begin{figure}[t!]
    \centering
        \includegraphics[width=\linewidth]{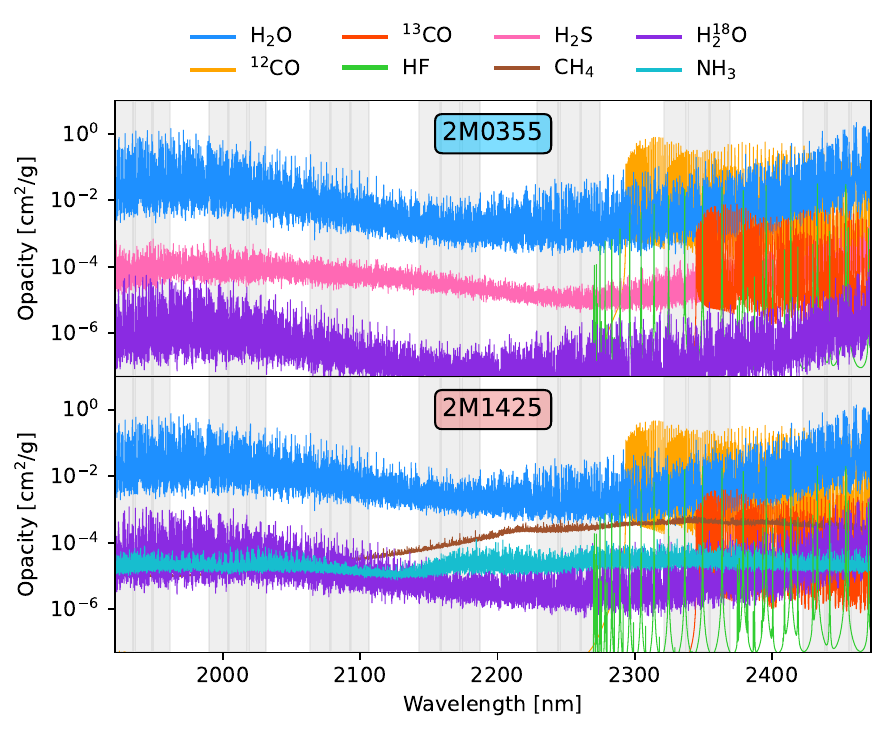}
     \caption{Most relevant opacity sources of 2M0355 (top) and 2M1425 (bottom), at the temperature present at the pressure where the emission contribution is maximal (1580\,K and 1850\,K, respectively). The opacities are scaled to the retrieved abundances (see Table~\ref{tab:free_params}). The vertically shaded regions show the wavelength coverage of the K2166-band.}
    \label{fig:opacities}
\end{figure}

Continuum opacities incorporated in the model spectra include collision-induced absorption from H$_2$--H$_2$ and H$_2$--He, as well as Rayleigh scattering of H$_2$ and He \citep{Dalgarno1962, Chan1965, Borysow1988}. To avoid introducing any biases by making assumptions about the thermochemical state of the atmosphere, we opt for the free chemistry approach, following previous studies (e.g., \citealt{deRegt2024, Picos2024}). In this approach, the volume mixing ratios (VMRs) of the retrieved chemical species are allowed to vary freely to produce the best fit to the data. This avoids biases and can enable a better fit in the case of disequilibrium chemistry in the atmosphere.

Our retrieval framework includes the atmospheric species listed in Table~\ref{tab:free_params}. We use the HITEMP line lists for $^{12}$CO, $^{13}$CO, C$^{18}$O, C$^{17}$O \citep{Li2015} and CH$_4$ \citep{Hargreaves2020}. The ExoMol line lists were used for H$_2$O (POKAZATEL; \citealt{Polyansky2018}), H$_2^{18}$O \citep{Polyansky2017}, NH$_3$ \citep{Coles2019}, HCN \citep{Harris2006, Barber2014}, HF \citep{Li2013, Coxon2015, Somogyi2021}, and H$_2$S \citep{Azzam2016, Chubb2018}. A selection of opacity sources are shown in Figure~\ref{fig:opacities} considering the retrieved abundances of 2M0355 and 2M1425 (see Table~\ref{tab:free_params} and
Section~\ref{sec:results}).

The VMRs of all species are kept constant with altitude to reduce the number of free parameters. As the atmospheres of brown dwarfs mainly consist of H$_2$ and He, we follow \cite{deRegt2024} and fix the helium abundance to VMR$_\text{He} =0.15$ and adjust the abundance of H$_2$ according to the VMRs of the retrieved species, to obtain a total VMR of unity. The VMRs, or number fractions $n_X$, of the included species $X$ are independently retrieved, from which further properties can be calculated. The CO isotopolog ratios for $^{13}$CO, C$^{17}$O, and C$^{18}$O in relation to $^{12}$CO, as well as the ratio of H$_2^{16}$O to H$_2^{18}$O, are calculated through the retrieved abundances of these molecules. The atmospheric C/O ratio is determined through the ratio of carbon and oxygen contained within all carbon- and oxygen-bearing species included in the retrieval. The atmospheric metallicity is approximated through the number fraction of carbon relative to hydrogen, scaled to the solar composition, with log$_{10}\left( \frac{n_\text{C}}{n_\text{H}} \right)_\odot$\,=\,$-3.54$ \citep{Asplund2021}:

\begin{equation}
    [\text{Fe/H}] \approx  [\text{C/H}] = \text{log}_{10}\left( \frac{n_\text{C}}{n_\text{H}} \right) - \text{log}_{10}\left( \frac{n_\text{C}}{n_\text{H}} \right)_\odot .
\end{equation}

For comparative purposes, as in \cite{deRegt2024}, we also run additional retrievals that assume chemical equilibrium. For this, we interpolate the abundances from equilibrium chemistry tables pre-computed with \texttt{FastChem} \citep{Stock2018, Kitzmann2024}, which depend on the pressure, temperature, metallicity, and C/O ratio, while assuming a solar N/O ratio. Therefore, instead of retrieving the individual abundances, the free parameters relating to the chemistry in this setup are the C/O ratio, metallicity, and base-10 logarithm of the isotope ratios $^{12}$C/$^{13}$C, $^{16}$O/$^{17}$O, and $^{16}$O/$^{18}$O. The chemical equilibrium abundances vary with altitude, in contrast to the free chemistry abundances, which are kept constant. 

\subsubsection{Pressure-temperature profile}

The temperature structure of an atmosphere determines its physical processes and chemical composition, and hence the observed spectral features. In isolated objects, such as the two brown dwarfs in our sample, the only source of heat originates from their interiors. Generally, the temperature at higher pressures (lower altitudes) is highest and decreases as one proceeds to lower pressures (higher altitudes). The modeling of exoplanet and brown dwarf atmospheres requires balancing physical consistency with computational efficiency. Parametrized models are a common approach, in which the temperature is defined at several pressure points and subsequently interpolated for the remaining atmospheric layers (e.g., \citealt{Line2015}). We implement a method based on \cite{Zhang2023}, who propose modeling the thermal profile through temperature gradients. The pressure-temperature ($P$--$T$) profile is modeled by dividing the atmosphere into 50 layers, which are evenly distributed in log$_{10}$-space in the pressure range of $10^2$--$10^{-6}$~bar. Free parameters in our retrieval include the temperature at the bottom of the atmosphere (denoted as $T_0$, at a pressure of $P_0$\,=\,$10^2$\,bar), and the temperature gradient $\nabla T_i$ at five equally spaced knots in pressure log$_{10}$-space, namely log$_{10}$($P$)\,=\,[-6, -4, -2, 0, 2]\,bar. Following \cite{Zhang2023}, the temperature gradient is defined as
\begin{equation}
    \nabla T_i = \frac{d \text{ln} T_i}{d \text{ln} P_i}.
\end{equation}

As $i$ increases, so does the altitude in the atmosphere, corresponding to a lower pressure. The retrieved gradients at the five knots are quadratically interpolated to obtain the gradients at all 50 atmospheric layers. These gradients are then used to calculate the temperature for each of the 50 layers:
\begin{equation}
    T_i = T_{i-1} \cdot \left( \frac{P_i}{P_{i-1}}  \right)^{\nabla T_i} .
\end{equation}

\subsection{Likelihood and covariance} \label{subsec:lncov}

For each model spectrum that is generated based on a sample of free parameters, a likelihood is evaluated to assess how well it matches the data. This likelihood is then used by \texttt{PyMultinest} to efficiently guide the exploration of the parameter space towards the best-fit model.
We implement a likelihood formalism based on the methods used in \cite{Ruffio2019}. First, the residuals $\vec{R}$ between the data spectrum $\vec{d}$ and the model spectrum $\vec{m}$ are calculated as
\begin{equation}
    \vec{R} = \vec{d} - \phi \vec{m}.
\end{equation}

The flux is scaled with the factor $\phi$ to account for instrument-induced differences between the spectral order-detector pairs and allow for a more accurate comparison between the data and the model. This factor is optimized for each model. For each order-detector pair, it is calculated as
\begin{equation}
    \phi = \frac{\vec{m}^T \vec{\Sigma_0}^{-1}\vec{d}}{\vec{m}^T \vec{\Sigma_0}^{-1}\vec{m}}.
\end{equation}

$\vec{\Sigma_0}$ represents the covariance matrix, which consists of the flux uncertainties and potential correlation between pixels. Following \cite{Ruffio2019}, for each order-detector pair, the log-likelihood is defined as
\begin{equation} \label{equ:lnL}
\begin{split}
\text{ln}\mathcal{L} = -\frac{1}{2} (&\text{ln}(|\vec{\Sigma}|) + \text{ln}(\vec{m}^T \vec{\Sigma}^{-1} \vec{m}) \\
& + (N_d - N_\phi + \alpha -1) \cdot \text{ln} \underbrace{(\vec{R}^T \vec{\Sigma}^{-1} \vec{R})}_{\chi_0^2})
\end{split}
\end{equation}

As in \cite{Ruffio2019}, we set $\alpha= 2$. $N_d$ is the number of valid datapoints (i.e., excluding masked pixels), and $N_\phi$ is the number of flux-scaling factors (in our case $N_\phi$\,=\,1, as we only have one per order-detector pair). The log-likelihood is calculated separately for each order-detector pair as shown in Equation~\ref{equ:lnL}, and the values are subsequently added together to obtain the total log-likelihood of the model. In this process, we also compute the uncertainty scaling factor $s^2$, with which we obtain the scaled covariance matrix $\vec{\Sigma}=s^2\vec{\Sigma}_0$ and a reduced chi-squared equal to unity: $\chi_\mathrm{red}^2=\vec{R}^T\vec{\Sigma}^{-1}\vec{R}/N_d=1$. The purpose of this scaling is to account for any over- or under-estimation of the flux-uncertainties, assuming a perfect model fit and uncorrelated noise \citep{Ruffio2019}. The uncertainties could be underestimated due to previous processing steps, such as the telluric removal, not being factored into the uncertainties adequately. For each order-detector pair, the uncertainty scaling factor is calculated as
\begin{equation} \label{equ:s2}
    s^2=\frac{1}{N_d}\vec{R}^T\vec{\Sigma}_0^{-1}\vec{R}.
\end{equation}

If there is no correlation between the individual pixels, the covariance matrix would consist only of the squared flux-uncertainties on its diagonal (i.e., $\Sigma_{0,ij} = \delta_{ij} \sigma_i^2$). However, uncertainties in the data can be correlated, which can occur through instrumental effects, telluric lines, other systematics, or imperfections in the line lists used for modeling. Not accounting for correlated noise can bias retrieved parameters and result in underestimated uncertainties \citep{Czekala2015, ZhangZ2021, deRegt2024}. The off-diagonal elements of the covariance matrix, representing the correlation between pixels, are modeled using Gaussian Processes (GP), which are then added to the uncorrelated component, i.e., the flux-uncertainties on the diagonal \citep{Czekala2015, Kawahara2022}. We introduce a Gaussian with the amplitude $a$, which scales the uncertainty, and a length-scale $\ell$, which sets the contribution of off-diagonal elements. The length-scale represents the extent of correlation in nanometers over which the correlation between two points decreases by a factor of 1/$e$. The covariance of pixels $i$ and $j$ with the separation $\lambda_i - \lambda_j$ are calculated as
\begin{equation}
    \Sigma_{0,ij} = \underbrace{\delta_{ij}\sigma_i^2}_{\substack{\text{uncorrelated}\\\text{component}}} + \underbrace{a^2 \sigma_{\mathrm{eff},ij}^2 \exp{\left(-\frac{(\lambda_i - \lambda_j)^2}{2\ell^2}\right)}}_\text{correlated component (GP)}. \label{eq:Sigma_0ij}
\end{equation}
Here, $\delta_{ij}$ is the Kronecker delta and $\sigma_i$ the flux-uncertainty of pixel $i$. For the effective uncertainty $\sigma_{\text{eff},ij}$, we use the median flux-uncertainty of each order-detector pair. The GP amplitude $a$ and length-scale $\ell$ are included as free parameters in our retrieval. Without masking, the covariance matrix for each order-detector pair would have a size of 2048\,$\times$\,2048, which is very computationally expensive. To improve efficiency, we only consider the off-diagonal elements until $(\lambda_i - \lambda_j) \leq 5\ell$, as the largest values are concentrated near the diagonal. As in \cite{deRegt2024}, we apply a banded Cholesky decomposition to the covariance matrix to compute the likelihood in Equation~\ref{equ:lnL}.

\subsection{Detection assessment} \label{subsec:detection}

After running the retrievals, the detection significance and opacity contribution of minor atmospheric species is quantified by performing a cross-correlation analysis and running additional retrievals. Following \cite{Zhang2021_SJ}, the cross-correlation is determined between a template of the selected species $\vec{m_{\text{only $X$}}}$ and the residuals, which constitute the difference between the observed spectrum $\vec{d}$ and the fiducial model excluding the selected species $\vec{m_{\text{w/o $X$}}}$. The template of the selected species is determined by subtracting the fiducial model excluding the selected species from the complete fiducial model including the selected species. The inverse covariance matrix $\vec{\Sigma}_0^{-1}$ is multiplied for weighting purposes. For each radial velocity shift $v$ of the template and each order-detector pair $ij$, the cross-correlation function (CCF) is calculated as
\begin{equation}
    \text{CCF}(v) =\sum_{ij} \frac{1}{s_{ij}^2} m_{ij,\text{only $X$}}(v)^T \Sigma_{0,ij}^{-1} (d_{ij}-m_{ij, \text{w/o $X$}}).
\end{equation}

The cross-correlation is computed for template velocity shifts between $\pm$500\,km/s in steps of 1\,km/s, relative to the object rest-frame. Per radial velocity shift, the cross-correlation coefficients of all order-detector pairs are summed. The noise is estimated through the standard deviation of the CCF, excluding the velocities within $\pm$100\,km/s. Finally, the signal-to-noise ratio (S/N) is determined by dividing the CCF by the noise, where a S/N\,$>$\,5 is considered a significant detection.

Furthermore, additional retrievals are performed for assessing the detection significance of some selected species. The additional retrievals have the same set-up as the fiducial model, except that we exclude each of the selected species $X$ at a time and determine the subsequent evidence (ln$\mathcal{Z}_\text{w/o $X$}$) of the best-fitting model without the selected species. This value is then compared to the evidence of the fiducial model (ln$\mathcal{Z}_0$), which includes the selected species, to determine how greatly the inclusion of the selected species is favored:
\begin{equation}
    \text{ln}B_{m, X}= \text{ln}\mathcal{Z}_0 - \text{ln}\mathcal{Z}_\text{w/o $X$}.
\end{equation}

The result is the logarithm of the Bayes factor ($B_m$), which is converted into a detection significance $\sigma$ of the selected molecule $X$ following \cite{Benneke2013}, with $\sigma$\,>\,3 constituting a significant detection.

\subsection{Validation with a synthetic spectrum} \label{subsec:test}

To validate the accuracy and performance of our retrieval framework, we run a test on a synthetic spectrum. We parametrize the $P$--$T$ profile in the same way as we do the forward modeling, namely by using temperature gradients at five pressure points, following a similar structure to the one retrieved for 2M0355 (see Figure~\ref{fig:summary}), and consider a surface gravity of log\,$g$\,=\,4.75. Furthermore, we include species with log$_{10}$ VMRs of $-3.00$ for H$_2$O and $^{12}$CO, $-5.00$ for $^{13}$CO and H$_2$S, and $-6.00$ for CH$_4$, HF and H$_2^{18}$O. We set the abundances of minor species (C$^{17}$O, C$^{18}$O, HCN, NH$_3$) to varying VMRs $\leq$\,$-$8 to test how the retrieval framework handles them. Considering their weak line strengths and very low abundances, these species are at levels that are not detectable. To mimic the properties of 2M0355, we apply a projected rotational velocity $v\,\text{sin}\,i$\,=\,4\,km/s. The same wavelength regions are masked as in the spectrum of 2M0355, where deep tellurics are found. We add Gaussian noise to the flux based on the mean uncertainty of 2M0355. The same priors as in Table~\ref{tab:free_params} are used. 

The results of the validation can be seen in Appendix~\ref{app:test}. Our retrieval framework is able to correctly reproduce the synthetic spectrum and $P$--$T$ profile. The abundances of the main constituents (with log$_{10}$ VMRs\,$\geq$\,$-$6.00) are retrieved with excellent accuracy, within $\pm\,0.08$ (in log$_{10}$-space), as well as the surface gravity at log\,$g$\,=\,4.74\,$\pm\,0.00$. Although the undetectable minor species should not be constrained at all, with cross-correlation yielding no signal for any of them (S/N\,$\leq$\,1), some of their posteriors exhibit a Gaussian-like shape, suggesting a low-level constraint at the 1\,$\sigma$ level. This artifact can emerge with parameters that do not affect the likelihood, causing \texttt{MultiNest} to continue drawing samples from the same region, resulting in a Gaussian distribution. The retrieved abundances of the undetectable species remain below $-$8 and do not affect the spectrum. Due to the excellent performance of our retrieval framework on the synthetic test spectrum, we expect it to reliably constrain the main atmospheric properties of 2M0355 and 2M1425.

\section{Results}\label{sec:results}

\begin{figure*}[ht!] 
     \centering
     \begin{subfigure}[b]{\textwidth}
         \centering
        \includegraphics[width=\textwidth]{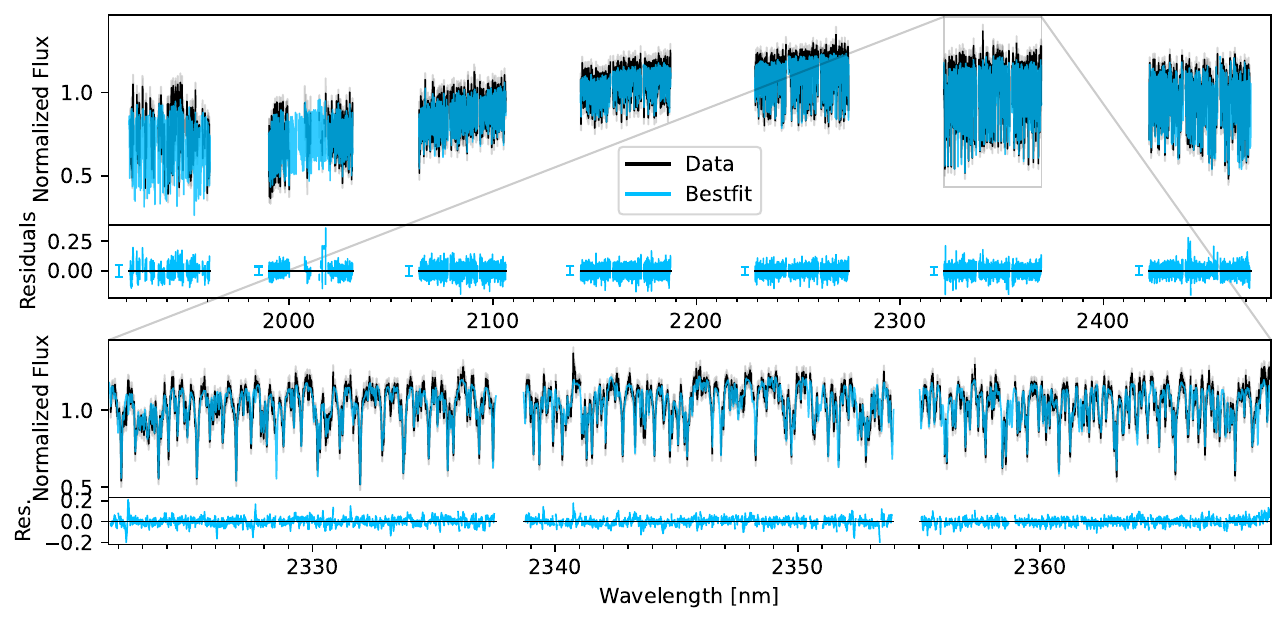}
     \end{subfigure}
     \begin{subfigure}[b]{\textwidth}
         \centering
    \includegraphics[width=\textwidth]{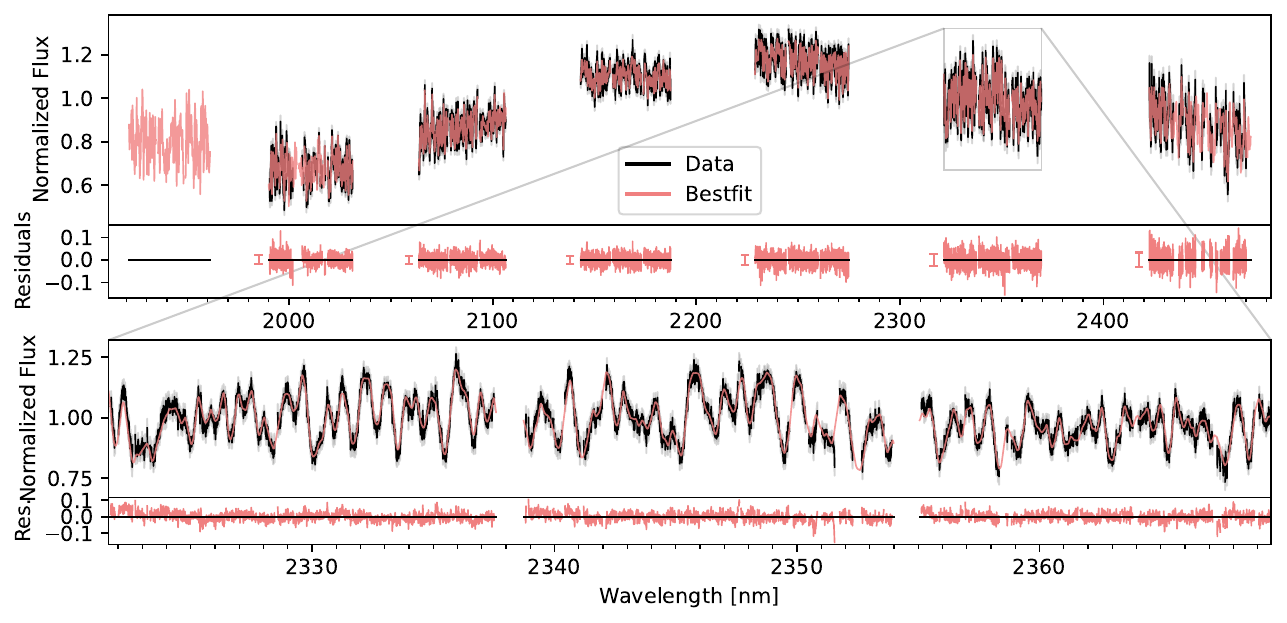}
     \end{subfigure}
     \caption{Best-fit spectra for 2M0355 (top) and 2M1425 (bottom). The observed spectrum is shown in black, the scaled 1\,$\sigma$ uncertainties in gray, and the best-fit model in blue and pink for 2M0355 and 2M1425, respectively. The residuals are in the panel underneath. The mean scaled uncertainties ($s_i$\,diag\,$\Sigma_{0,i}$) for each spectral order are shown as errorbars next to the residuals.}
     \label{fig:bestfit}
\end{figure*}

The retrieved free parameters, including their errors, are listed in Table~\ref{tab:free_params} for 2M0355 and 2M1425, for both free and equilibrium chemistry. The quality of the fits, as well as a comparison between the free and equilibrium chemistry retrievals, are assessed in the last section of the table. We note that the small errors on the retrieved parameters may not be realistic, particularly considering that too narrow posteriors are a commonly reported issue of \texttt{MultiNest} retrievals (e.g., \citealt{Buchner2016, Ardevol2022}). The following sections focus on the results of the free chemistry setup, unless stated otherwise.

\subsection{Best-fit spectra} \label{subsec:bestfit}

Figure~\ref{fig:bestfit} shows the best-fitting spectra for 2M0355 (top, in blue) and 2M1425 (bottom, in pink) across the entire observed wavelength region, including the residuals below each spectral order and an inset zoomed in on a portion of the spectrum for a clearer view of the lines. Appendix~\ref{app:bestfit} provides a more extensive view of each spectral order. As seen from the close overlap of the data and model spectra, as well as the residuals, which mainly lie within the (scaled) uncertainties (shown in gray around the observed spectrum and as errorbars next to the residuals), both model spectra provide an accurate fit to the observed spectral features. Small imperfections can be observed in the residuals, such as sharp increases and subsequent decreases (or vice-versa), which may be the result of suboptimal line positions or inaccuracies in the wavelength solution. The most relevant opacity sources contributing to the spectra of 2M0355 and 2M1425 are shown in Figure~\ref{fig:opacities}.

For both objects, we find that the free and equilibrium chemistry retrievals produce a similarly good fit to the spectrum according to the $\chi^2$ (see Table~\ref{tab:free_params}), with the free chemistry being only marginally better. The differences in log-likelihood (ln\,$\mathcal{L}$ from Equ.~\ref{equ:lnL}) of the respective best-fit models also indicate that the free chemistry approach produces a better fit to the data of both objects. However, the Bayes factor ln\,$B_m$, which compares the global log-evidence of the free and equilibrium chemistry, favors the equilibrium model. This is because there are fewer free parameters required for the equilibrium retrieval, while producing a similarly good fit. The evidence reflects model complexity through the integration over the entire parameter space, hence disfavoring models with more parameters, unless they significantly improve the likelihood. While the equilibrium chemistry setup requires fewer parameters to fit the data, the free chemistry setup offers the advantage of quantifying the abundances of individual species. In this regard, both models can be considered valid, each with its own advantages and drawbacks. 

\subsection{Velocities and gravity}

\begin{figure*}[ht!]
    \centering
    \includegraphics[width=\linewidth]{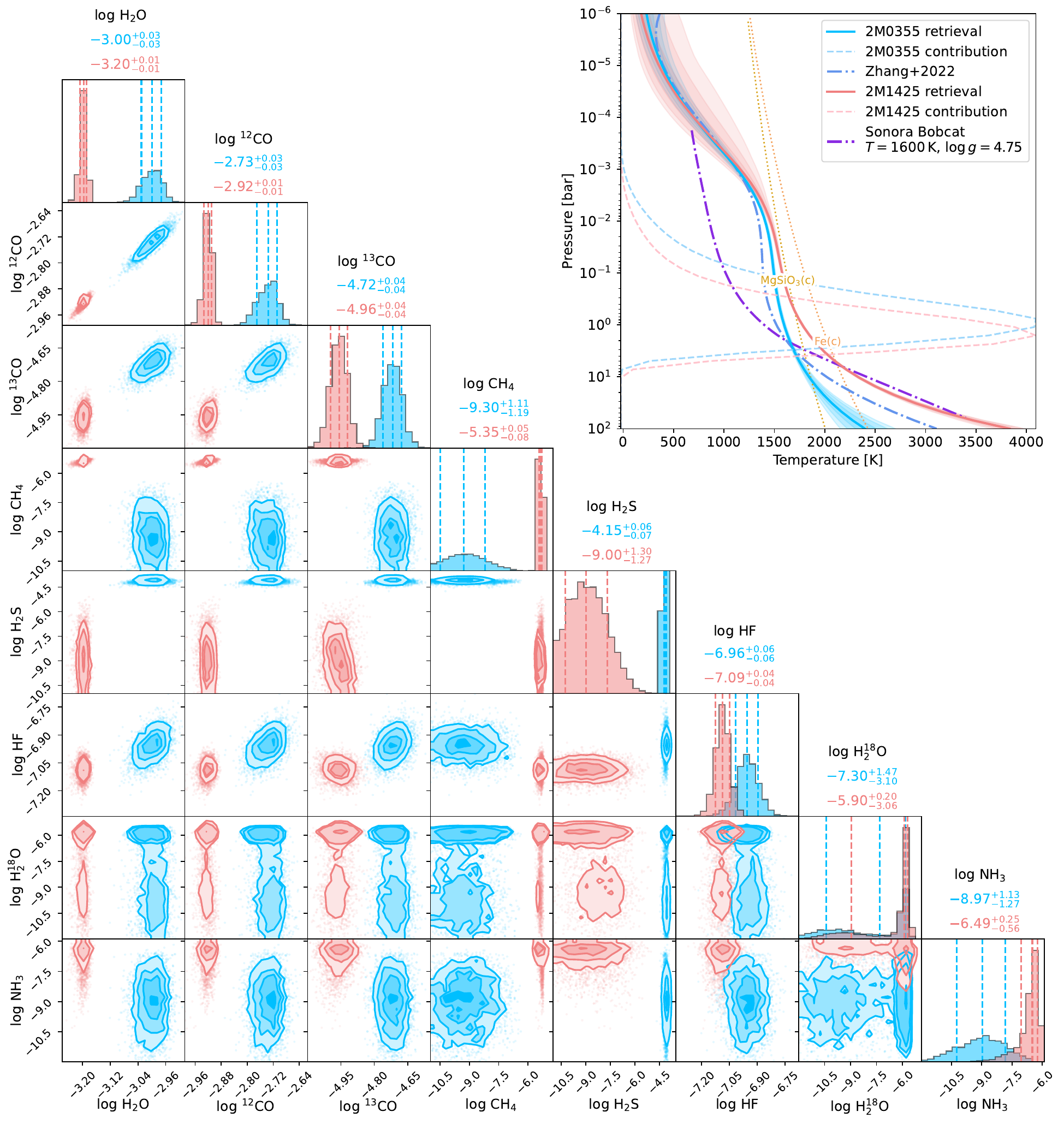}
    \caption{Retrieval results showing the posterior distributions of atmospheric species and the thermal structure of 2M0355 (blue) and 2M1425 (pink). Upper right panel: $P$--$T$ profiles (solid lines) and their 68, 95, and 99.7\% confidence intervals (shaded regions around them). The integrated emission contribution function for both objects are shown in their respective colors as vertical intensity functions on the pressure axis. A solar-metallicity Sonora Bobcat $P$--$T$ profile ($T_\text{eff}$\,=\,1600\,K, log\,$g$\,=\,4.75) is shown in purple for comparison \citep{Marley2021}. The $P$--$T$ profile retrieved for 2M0355 by \cite{Zhang2022} is shown as a blue dashed-dotted line. The condensation curves of MgSiO$_3$ and Fe \citep{Visscher2010} are included as dotted lines. Lower left panels: Cornerplot displaying the retrieved posterior distributions of selected chemical species of 2M0355 and 2M1425. The median retrieved values, along with the 16$^\text{th}$ and 84$^\text{th}$ percentile, are shown as vertical dashed lines.}
    \label{fig:summary}
\end{figure*}

Our retrieved radial velocities are similar to the literature values (see Table~\ref{tab:properties}), with differences being $\Delta v_\text{rad}$\,=\,0.17\,km/s for 2M1425 and $\Delta v_\text{rad}$\,=\,1.33\,km/s for 2M0355. The visibly most notable difference between the two spectra are the effects of the rotational broadening, making the absorption lines of 2M1425 shallower and significantly broadened, while those of 2M0355 are much deeper and easier to separate. As expected from this, the retrieved projected rotational velocity $v\,\text{sin}\,i$ is significantly higher for 2M1425 ($\sim$32\,km/s) than for 2M0355 ($\sim$3\,km/s). This is consistent with the literature value for 2M1425 \citep{Blake2010}, while for 2M0355 it is consistent with the smaller value (<\,4\,km/s) found by \cite{Zhang2021_BD} in contrast to $14.7\,\pm\,2.3$\,km/s reported by \cite{Bryan2018}. This discrepancy is likely a result of the lower resolution $\mathcal{R}$\,=\,$25.000$ in \cite{Bryan2018}, which can only resolve down to 12\,km/s, while our data of $\mathcal{R}$\,=\,$90.000$ can resolve up to 3\,km/s.

The differences in $v\,\text{sin}\,i$ could reflect simply a difference in viewing angle, i.e., our line of sight may be aligned with the rotational axis of 2M0355, making it appear to rotate slower compared to 2M1425. However, \cite{Vos2022} report a variability with a period of $\mathcal{P}$\,=\,9.53\,h for 2M0355, while \cite{Vos2019} find a low-amplitude trend with a period of $\mathcal{P}$\,$\gtrsim$\,2.5\,h for 2M1425, which would imply a significantly higher rotational velocity if 2M1425's period is close to the reported lower limit. Consequently, 2M0355's lower $v\,\text{sin}\,i$ compared to 2M1425 indeed appears to be due to slower rotation. However, a projection effect to some extent is still possible and likely, as a period of 9 hours implies a relatively fast rotation, considering that young brown dwarfs can exhibit rotation periods between a few hours to over a day \citep{Moore2019}.

The rotational velocity can reflect different evolutionary stages. As brown dwarfs age and cool, they contract and therefore increasingly rotate faster due to the conservation of angular momentum \citep{Bouvier2014, Bryan2020}. As both objects have similar masses (see Table~\ref{tab:properties}), they can be expected to evolve at a similar rate. A contraction with aging would also increase the surface gravity log\,$g$. Indeed, the retrieved log\,$g$ of the faster rotator 2M1425 ($\sim$\,4.98\,$\pm$\,0.01) is significantly higher than for 2M0355 ($\sim$\,4.75\,$\pm$\,0.03), even in the equilibrium chemistry retrieval. However, age differences within young moving groups are expected to be within $\sim$\,10\,Myr, making large differences in evolutionary stages unlikely \citep{Goldman2018, Gagne2023}. Instead, the rotational velocity is more likely to be affected by underlying physical differences resulting from different initial conditions in their birth environment, accretion history, and dynamical interactions. Due to these factors, rotational velocities of brown dwarfs can vary significantly even within coeval populations \citep{Rodriguez2009, Scholz2015}. 

Interestingly, our retrieved surface gravity of log\,$g$\,$\sim$\,4.75\,$\pm$\,0.03 (and log\,$g$\,$\sim$\,4.73\,$\pm$\,0.04 assuming equilibrium chemistry) for 2M0355 is significantly higher than the one retrieved by \cite{Zhang2021_BD}, being 4.32\,$\pm$\,0.15. This discrepancy could arise from the fact that K-band observations lack reliable features sensitive to gravity \citep{Zhang2021_BD}. Furthermore, it is important to note that using a narrow wavelength range can lead to inaccurate retrievals of the surface gravity, as well as other fundamental parameters such as the mass and radius \citep{Burningham2021}. Although the surface gravity was retrieved correctly for the synthetic test spectrum (see Section~\ref{subsec:test}), we caution that this is for the ideal case.

\subsection{Thermal profile and clouds}

Figure~\ref{fig:summary} shows a summary and comparison of the retrieval results for 2M0355 (blue) and 2M1425 (pink). The posterior distributions of other parameters can be found in Appendix~\ref{app:cornerplot_rest}. The retrieved $P$--$T$ profiles of both objects, shown in the top right panel, are very similar at low pressures, but diverge towards higher pressures, indicating that 2M1425 is somewhat hotter than 2M0355. The emission contribution function, integrated over all wavelengths, indicates that the observed spectral features for both objects mainly stem from pressures around 1\,bar, supported by the narrower uncertainties of the temperature profile in this region of the atmosphere. Their temperatures in the photosphere, as illustrated by their emission contribution functions, differ by about 50--200\,K. This agrees with the effective temperatures derived from their spectral energy distributions, which also suggest that 2M1425 ($T_\text{eff}$\,=\,1535\,$\pm$\,55\,K) may be slightly hotter than 2M0355 ($T_\text{eff}$\,=\,1478\,$\pm$\,57\,K), though their temperatures overlap within the errors \citep{Filippazzo2015}. The higher surface gravity of 2M1425 is also reflected in the somewhat higher pressures of its emission contribution function. Overall, their temperature profiles deviate considerably from the shape
of the Sonora Bobcat $P$--$T$ profile \citep{Marley2021} shown in purple, chosen due to its temperature and surface gravity ($T_\text{eff}$\,=\,1600\,K, log\,$g$\,=\,4.75) similar to the literature values of our two objects. Furthermore, the retrieved $P$--$T$ profile of 2M0355 is similar to the one obtained by \cite{Zhang2022} in the science verification observations of CRIRES+, who parametrized the $P$--$T$ profile using four temperature knots.

For both objects and in both chemistry setups, we find no clear evidence for clouds using our gray cloud model, as suggested by the negligible cloud opacity and unconstrained cloud properties. This is in agreement with the free retrieval (without enforcing clouds) of 2M0355 by \cite{Zhang2021_BD}, and similar to other brown dwarfs of the SupJup Survey \citep{deRegt2024, Mulder2025}. As clouds alter the slope of the continuum, the narrow wavelength range used for high-resolution spectroscopy significantly complicates their retrieval. The lack of evidence for clouds in most retrieval analyses using high resolution spectra is therefore unsurprising and even expected \citep{Xuan2022, Landman2024}. For example, retrievals by \cite{Burningham2021} have demonstrated that using only narrow bands such as J, H, or K, can lead to degeneracies and biases in the retrieved thermal structures, fundamental parameters (log\,$g$, mass, and radius) and cloud properties.

Compared to 2M1425, 2M0355 has been reported to have a significantly redder continuum \citep{Gagne2015}, which could be the result of clouds or a $P$--$T$ profile with a reduced temperature gradient \citep{Tremblin2015}. Figure~\ref{fig:summary} shows that 2M0355 is indeed more isothermal in the photosphere than 2M1425, though this does not necessarily rule out the need for clouds. It is possible that our cloud model is not adequately capturing the reddening, which would force the $P$--$T$ profile to become more isothermal to compensate. The condensation curves in the $P$--$T$ profile plot in Figure~\ref{fig:summary} indicate that MgSiO$_3$ may form clouds in 2M1425 at pressures probed by the observations. However, the cloud opacity could be negligible or inadequately reproduced by our gray cloud model. Furthermore, the K-band is expected to be less sensitive to clouds than the J- and H-band with shorter wavelengths that probe deeper in the atmosphere \citep{Landman2023}. More likely for both objects, however, is that the cloud properties are simply not retrievable due to the narrow wavelength range. Larger wavelength ranges, such as data from JWST, are better suited for detecting clouds \citep{Miles2023}.

\subsection{Chemical composition}

\begin{table}[t!]
    \centering
    \caption{Detection assessment of various atmospheric species.}
    \begin{tabular}{l|rr|rr}
    \hline
    \hline
    & \multicolumn{2}{c}{2M0355} & \multicolumn{2}{c}{2M1425} \\
    Species & S/N & $\sigma$ & S/N & $\sigma$ \\
         \hline
    $^{13}$CO & 12.4 & 13.4 & 3.5 & 8.0 \\
HF & 16.9 & 11.6 & 5.5 & 15.8 \\
H$_2$S & 6.2 & 4.6 & 0.2 & 2.3 \\
H$_2^{18}$O & 2.6 & 1.1 & 1.5 & 3.0 \\
CH$_4$ & -0.7 & 2.2 & 5.5 & 5.5 \\
NH$_3$ & -1.0 & -1.4 & 1.6 & 3.0 \\
        \hline
    \end{tabular}
    \tablefoot{The S/N was determined through cross-correlation, while the significance $\sigma$ was determined through additional retrievals without the species in question (see Section~\ref{subsec:detection}).}
    \label{tab:evidence}
\end{table}

\begin{figure}[t!]
    \centering
    \includegraphics[width=\linewidth]{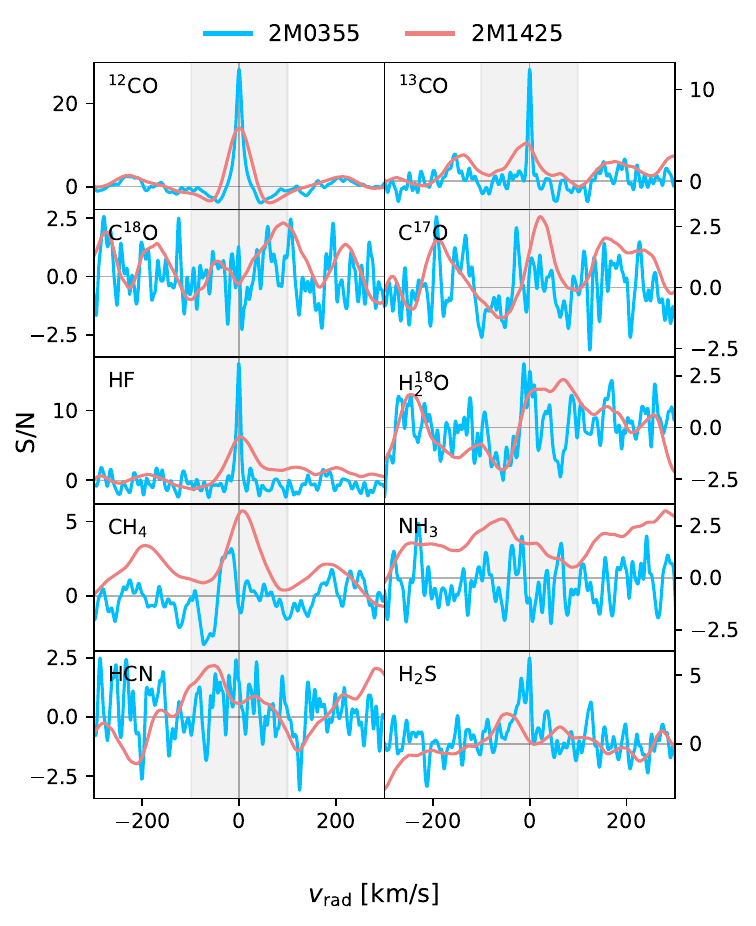}
    \caption{Cross-correlation functions of atmospheric species in 2M0355 (blue) and 2M1425 (pink). The noise was calculated from the standard deviation outside the shaded region ($\pm$\,100\,km/s).}
    \label{fig:CCFs}
\end{figure}

The retrieved abundances of the two brown dwarfs shown in Figure~\ref{fig:summary} and Table~\ref{tab:free_params} exhibit many similarities, in particular regarding the main (besides H$_2$ and He) atmospheric constituents H$_2$O, $^{12}$CO, and $^{13}$CO. HF is retrieved at a similar VMR in both objects. The detection of HF has also been reported for other brown dwarfs in the SupJup sample \citep{Picos2024, Zhang2024, Mulder2025, deRegt2025}. Due to its prominent absorption lines within $2.3$--$2.5\,\mu$m (see Figure~\ref{fig:opacities}), the K-band is particularly well suited for its detection, even at low abundances. Despite the different retrieved medians of H$_2^{18}$O between 2M0355 and 2M1425, their posterior distributions do exhibit clear peaks at similar abundances (log$_{10}$VMR\,$\sim$\,-6). Due to the much larger uncertainty towards lower abundances, we can consider a log$_{10}$VMR\,$\sim$\,-6 as an upper limit of H$_2^{18}$O in both objects.
The most notable compositional difference between the two objects is the detection of H$_2$S in 2M0355, while we retrieve CH$_4$ and NH$_3$ in 2M1425. So far, H$_2$S, which is the main sulfur-bearing species in substellar atmospheres \citep{Visscher2006}, has primarily been observed in cooler brown dwarfs of T and Y spectral types \citep{Tannock2022, Hood2023, Lew2024}.

As shown in Figure~\ref{fig:summary} and Table~\ref{tab:free_params}, we find $^{13}$CO, HF, and potential traces of H$_2^{18}$O in both 2M0355 and 2M1425, while H$_2$S is only retrieved in 2M0355, in contrast to CH$_4$ and NH$_3$ only retrieved in 2M1425. For each species, the cross-correlation S/N and detection significance $\sigma$ is assessed as described in Section~\ref{subsec:detection}. The results are listed in Table~\ref{tab:evidence} for both brown dwarfs. As evidenced by both the S/N and $\sigma$, the detection of $^{13}$CO, HF, and H$_2$S is robust in 2M0355. For 2M1425, the detection significance for $^{13}$CO, HF, and CH$_4$ is also robust, while it is more tentative for NH$_3$. Similar to \cite{Zhang2022}, who report only a tentative constraint on $^{18}$O in 2M0355 using CRIRES+, we also find only hints of H$_2^{18}$O in both objects.

Figure~\ref{fig:CCFs} shows the cross-correlation S/N of the atmospheric species in both objects. As the cross-correlation was performed in the rest-frame of the observations, the signal of detected species peaks at velocities of around 0\,km/s, demonstrating the detection of $^{12}$CO, $^{13}$CO, and HF in both objects, as well as H$_2$S in 2M0355 and CH$_4$ in 2M1425. The CCFs of 2M1425 are broader and have a lower S/N peak, as expected due to its larger $v\,\text{sin}\,i$, which smears out the signal of the spectral features. Interestingly, cross-correlating with NH$_3$ yields no detection for 2M1425, despite being retrieved with a significance of 3.0\,$\sigma$. As seen in Figure~\ref{fig:opacities}, it appears that NH$_3$ may mainly be contributing to the continuum rather than producing distinguishable features in the spectrum.

\begin{figure}[t!]
    \centering
    \includegraphics[width=\linewidth]{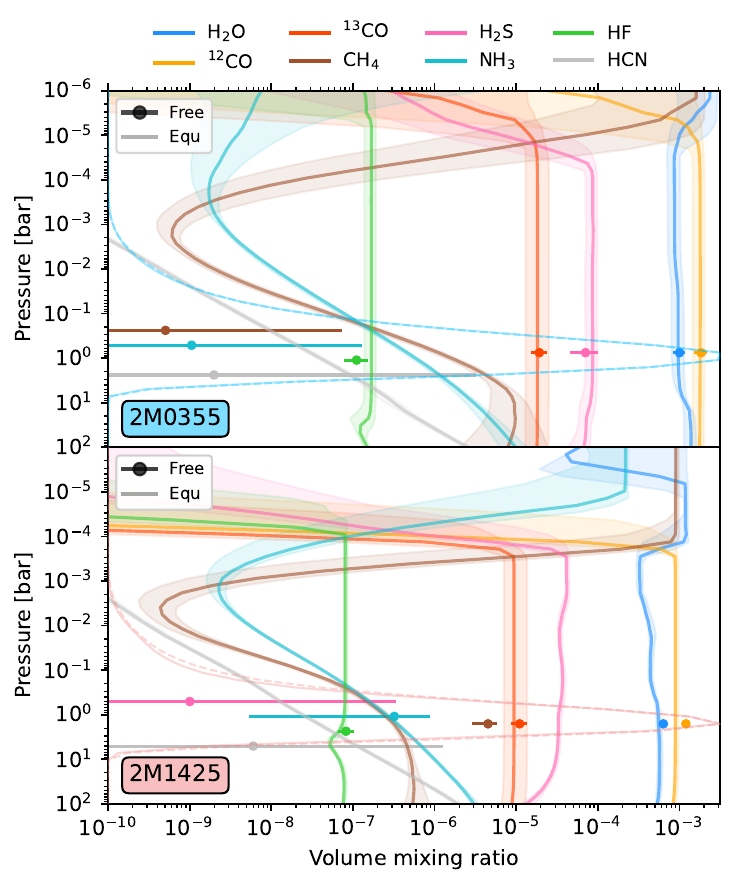}
    \caption{Volume mixing ratios (VMRs) in 2M0355 (top) and 2M1425 (bottom) of the main atmospheric species retrieved using free versus equilibrium chemistry. For clarity, the free chemistry abundances are only shown as markers with errorbars, while the equilibrium chemistry abundances, which vary throughout the atmosphere, are shown as solid lines. The errorbars and shaded envelopes show the retrieved 95\% confidence interval. The integrated emission contribution function (dotted: free, solid: equilibrium ) is shown on the pressure axis for reference.}
    \label{fig:VMR}
\end{figure}

As described in Section~\ref{subsec:bestfit}, both free and equilibrium chemistry models provide a similarly good fit to the data of both objects, indicating that the main contributors to the spectrum are in chemical equilibrium. However, not all abundances in the free and equilibrium models match, as seen in Figure~\ref{fig:VMR}. For one, the equilibrium chemistry models indicate that H$_2$S is expected for both objects at abundances up to a log$_{10}$VMR$\sim\,$-4 to -5. This is in perfect agreement with the free chemistry model of 2M0355, but no H$_2$S is retrieved in 2M1425. Furthermore, CH$_4$ is expected in the photosphere (around 1\,bar) at a log$_{10}$VMR$\sim\,$-6, with increasing abundances in the upper atmosphere, where the lower temperatures cause CO to be converted to CH$_4$. However, no CH$_4$ is retrieved for 2M0355, while we find somewhat a higher abundance for 2M1425. The retrieval of NH$_3$ in 2M1425 appears to agree perfectly with chemical equilibrium, though its large errors make this result only tentative. The deviations of CH$_4$ and H$_2$S from equilibrium could be due to disequilibrium chemistry processes, such as vertical mixing \citep{Mukherjee2022}, or also latitudinal variations \citep{Lee2024}. Different viewing angles, and hence observed latitudes, may to some extent be responsible for their different $v\,\text{sin}\,i$, with 2M0355 possibly being viewed pole-on and 2M1425 viewed from the side.

Minor atmospheric species (C$^{17}$O, C$^{18}$O, HCN, NH$_3$) are unconstrained, with upper limits at abundances of $\lessapprox$\,$-$8, except for NH$_3$ in 2M1425. The remaining abundances with a log$_{10}$VMR\,$\lessapprox$\,$-$8 are compatible with a non-detection. Despite appearing to be Gaussian, this may only be an artifact from \texttt{MultiNest}. The posteriors are very broad and cover several orders of magnitude. As investigated in Section~\ref{subsec:test} on a synthetic test spectrum, log$_{10}$VMRs\,$\lessapprox$\,$-$8 for species with weak spectral features cannot be accurately retrieved and vary with each independent run of the retrieval, despite Gaussian posterior distributions. The actual log$_{10}$VMRs may be any value below $-8$, as it is not possible to reliably constrain them at such low abundances.

\subsection{Abundance ratios and metallicity}

\begin{figure}
    \centering
    \includegraphics[width=\linewidth]{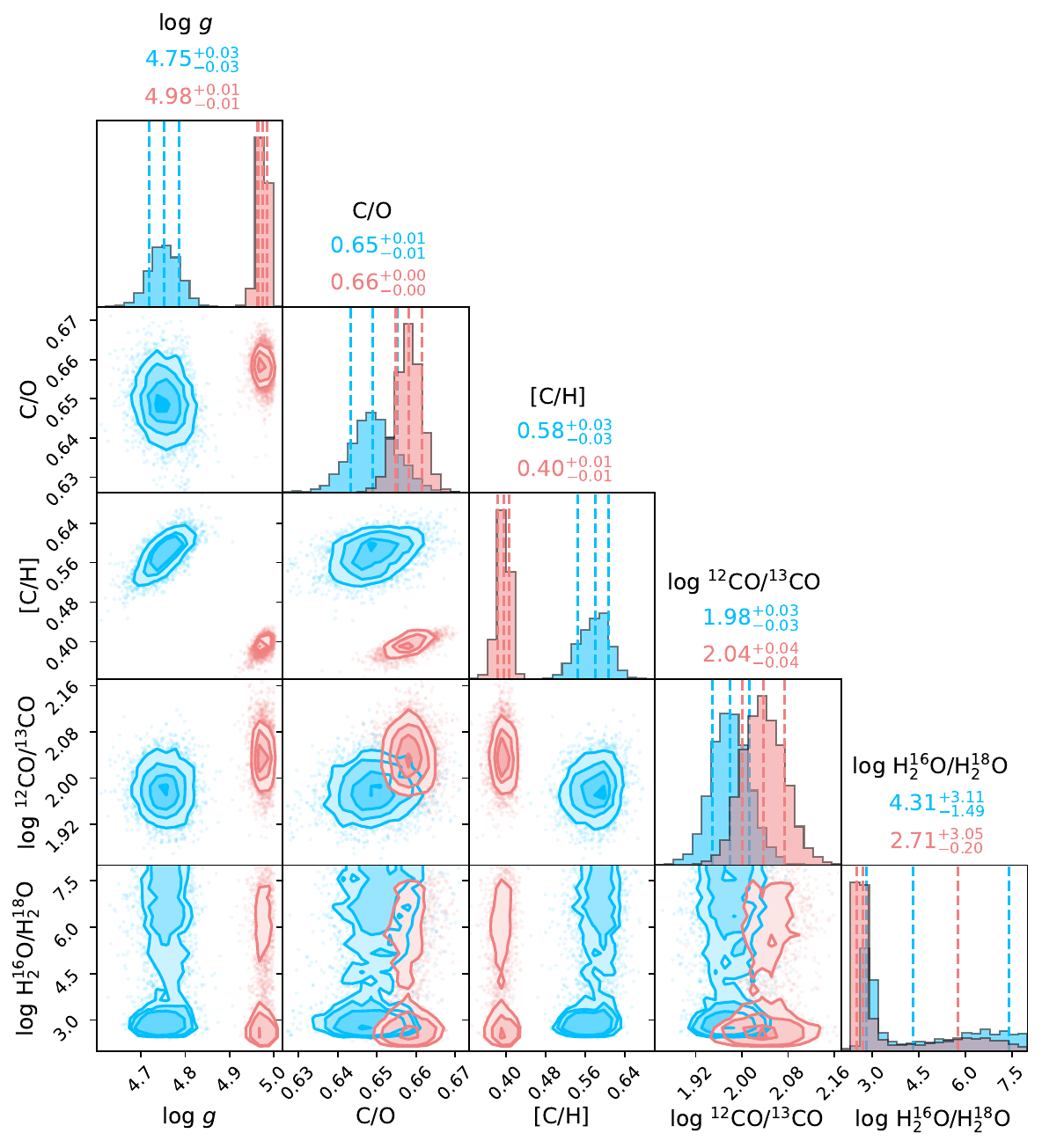}
    \caption{Posterior distributions of the surface gravity log\,$g$; C/O ratio; metallicity, approximated by the carbon-to-hydrogen ratio [C/H]; $^{12}$CO/$^{13}$CO; and H$_2^{16}$O/H$_2^{18}$O for 2M0355 (blue) and 2M1425 (pink). The C/O ratio and metallicity were not retrieved as free parameters, but determined from the retrieved abundances. An equivalent plot comparing the free and equilibrium chemistry setup can be found in Appendix~\ref{app:free_equ}.}
    \label{fig:ratios_logg}
\end{figure}

The isotopolog ratios, C/O ratio, and metallicity for both 2M0355 and 2M1425 shown in Table~\ref{tab:free_params} and Figure~\ref{fig:ratios_logg} are derived from the retrieved abundances. The metallicity for the free chemistry setup is estimated through the carbon-to-hydrogen ratio [C/H]. A comparison between the free and equilibrium chemistry setup can be found in Appendix~\ref{app:free_equ}.

We find that the C/O ratios of both objects are almost identical. Our value for 2M0355 of C/O\,$\sim$\,0.65$\substack{+0.01 \\ -0.01}$ is somewhat higher than the value reported by \cite{Zhang2021_BD} of C/O\,$\sim$\,0.55$\substack{+0.02 \\ -0.01}$, which may arise due to their retrievals assuming chemical equilibrium. Our equilibrium chemistry retrievals also result in a lower C/O ratio for both objects, being $\sim$\,0.52$\substack{+0.00 \\ -0.00}$ for 2M0355 and $\sim$\,0.57$\substack{+0.00 \\ -0.01}$ for 2M1425. This discrepancy could be a result of oxygen ending up in refractory condensates, such as MgSiO$_3$ clouds (e.g., \citealt{Line2021}). While we only measure C/O from the gaseous constituents using the free chemistry retrievals, the C/O of the equilibrium chemistry retrievals considers the entire atmospheric C/O, even that contained within clouds, naturally resulting in a lower C/O. Although our retrieval does not constrain the presence of clouds, MgSiO$_3$ clouds cannot be ruled out for 2M0355. For lower C/O ratios, the condensation curve of MgSiO$_3$ (see Figure~\ref{fig:summary}) would be shifted to slightly lower temperatures (see Fig.~3 in \citealt{Zhang2021_BD}), allowing it to intersect with the $P$--$T$ profile of 2M0355 within the emission contribution range.

Our results indicate that the metallicity of 2M1425 ([C/H]\,=\,0.40\,$\pm$\,0.01) is notably lower than that of 2M0355 ([C/H]\,=\,0.58\,$\pm$\,0.03), as it has a lower $^{12}$CO abundance, the main atmospheric species determining our calculation of [C/H]. Assuming chemical equilibrium, we obtain a similar metallicity for 2M0355, but an even lower one for 2M1425 at [Fe/H]\,$\sim\,$0.26\,$\pm$\,0.01. \cite{Zhang2021_BD} report a solar metallicity for 2M0355 ([Fe/H]\,=\,0.05$\substack{+0.11 \\ -0.10}$), which is in line with the metallicity of low-mass stars in the AB Dor moving group ($\sim$\,0.1, \citealt{Biazzo2012}). Furthermore, as shown in Figure~\ref{fig:ratios_logg}, we find a correlation between the metallicity and surface gravity, which has also been reported in other studies (e.g., \citealt{Zhang2021_BD, Picos2024}). It is particularly pronounced for 2M0355, for which we calculate a correlation coefficient of $r$\,=\,0.70 (with $r$\,=\,1 implying a perfect positive correlation and $r$\,=\,0 implying no correlation), while for 2M1425 it is weaker with $r$\,=\,0.47. We caution that this correlation reduces the confidence in the retrieved metallicity and log\,$g$.

Both objects exhibit comparable CO isotopolog ratios. While C$^{17}$O and C$^{18}$O are not detected, the retrieved ratios of $^{12}$CO/$^{13}$CO\,$\sim$\,100 are almost identical for both objects, being $95.5\substack{+6.8 \\ -6.4}$ for 2M0355 and $109.6\substack{+10.6 \\ -9.6}$ for 2M1425. This reflects a depletion of $^{13}$CO with respect to the ISM ($^{12}$C/$^{13}$C$_\text{ISM}$\,$=68 \pm 15$, \citealt{Milam2005}), which has also been observed in other brown dwarf atmospheres \citep{Costes2024, deRegt2024, Picos2024}, although with a large scatter, with some being compatible with the ISM within the observed range. As both of these brown dwarfs are part of the same moving group, their equivalent isotopolog ratios could indicate that the ISM had a ratio of $^{12}$CO/$^{13}$CO\,$\sim$\,100 at the time and in the region of their formation. In this regard, if any substellar companions or exoplanets within the AB Dor moving group exhibit deviating values, this could hint at a truly distinct formation pathway.

The relatively high $^{12}$C/$^{13}$C ratios compared to the present-day local ISM provide valuable measurements to test predictions of chemical evolution models \citep{Romano2022}. Moreover, the analysis of the entire sample of SupJup objects could shed more light on possible trends. Our results for 2M0355 also align with the $^{12}$CO/$^{13}$CO ratios derived for 2M0355 with Keck/NIRSPEC data ($97\substack{+25 \\ -18}$, \citealt{Zhang2021_BD}) and with VLT/CRIRES+ ($108 \pm 10$, \citealt{Zhang2022}). 

Compared to \cite{Zhang2022}, who report a tentative constraint on the oxygen isotope ratio of $^{16}$O/$^{18}$O\,=\,1489$\substack{+1027\\ -426}$ in 2M0355 using CRIRES+, we retrieve similarly poorly constrained ratio considering $^{18}$O from water, while the $^{18}$O from carbon monoxide yields even higher $^{16}$O/$^{18}$O ratios. Due to the narrower wavelength range of CO spectral features in the K-band compared to those of H$_2$O, it is unsurprising that $^{18}$O is more readily detectable in H$_2$O. The resulting H$_2^{16}$O/H$_2^{18}$O ratios, as seen in Figure~\ref{fig:ratios_logg}, exhibit log$_{10}$ peaks between 2.7 and 3 (i.e. 500--1000), which is higher than the tentative ratios obtained by \cite{Picos2024} for two isolated brown dwarfs (100--300). Due to the much larger uncertainty towards higher ratios, this result is tentative and more of a lower limit. These peaks are also seen in the $^{16}$O/$^{18}$O posterior from the equilibrium chemistry setup (see Appendix~\ref{app:free_equ}), though far less pronounced for 2M1425.

\subsection{Correlated noise}

Our results indicate a negligibly low degree of correlated noise in the spectra, whereas it appears to be slightly higher for 2M0355. Both the free and equilibrium chemistry setup retrieve the same values for the GP parameters. We find a GP amplitude $a$\,=\,1.9\,$\pm\,0.01$ and length-scale $\ell$\,=\,0.018\,$\pm\,0.00$\,nm\,$\approx$\,2.5\, pixels for 2M0355, while $a$\,=\,1.4\,$\pm\,0.01$ and $\ell$\,=\,0.012\,$\pm\,0.00$\,nm\,$\approx$\,1.8\,pixels for 2M1425 (see Section~\ref{subsec:lncov}). This is in contrast to \cite{Picos2024}, who find that the degree of correlated noise is higher for objects with a higher $v\,\text{sin}\,i$, while in our case, the slower rotator 2M0355 exhibits more correlated noise. Interestingly, the retrieval of our synthetic test spectrum yields similar values of $a$\,=\,3.6\,$\pm\,0.01$ and $\ell$\,=\,0.023\,$\pm\,0.00$\,nm\,$\approx$\,3.6\,pixels, despite having no correlated noise added to it. This shows that our framework may not be able to distinguish between uncorrelated noise and very low levels of correlated noise. It could be a similar artifact as the Gaussian posteriors of undetectable species, where \texttt{MultiNest} stops exploring parameters that do not change the likelihood (see Section~\ref{subsec:test}). Therefore, the retrieved low-level correlated noise for 2M0355 and 2M1425 can be considered be consistent with uncorrelated noise.

\section{Conclusion} \label{sec:concl}

In conclusion, we performed atmospheric retrievals on high-resolution CRIRES+ spectroscopy of two brown dwarfs of similar spectral type in the AB Doradus Moving Group, namely 2MASS J03552337+1133437 (2M0355) and 2MASS J14252798-3650229 (2M1425), revealing their composition and thermal structure. The retrieved abundances of both objects are similar regarding the main atmospheric species, and their $P$--$T$ profiles also exhibit similar shapes. We find robust evidence of $^{13}$CO and HF in both objects, with only weak hints of H$_2^{18}$O. Both objects have a similar isotopolog ratio of $^{12}$CO/$^{13}$CO\,$\sim$\,100. The brown dwarf with the higher $v\,\text{sin}\,i$, 2M1425, also exhibits a larger surface gravity, is slightly hotter, and shows clear evidence of CH$_4$ in its atmosphere with a non-detection of H$_2$S. In contrast, in the other brown dwarf 2M0355, CH$_4$ is depleted and H$_2$S is robustly detected instead. While the abundances of the main contributors to the spectrum, namely H$_2$O, $^{12}$CO, $^{13}$CO, and HF, align with our chemical equilibrium retrieval,  CH$_4$ (for 2M0355) and H$_2$S (for 2M1425) deviate from equilibrium. These results suggest that their atmospheres are mainly governed by chemical equilibrium, with possibly some disequilibrium processes affecting CH$_4$ and H$_2$S. Future analyses of the remaining sample of isolated brown dwarfs and substellar companions observed as part of the ESO SupJup Survey will provide further detailed insights into the atmospheres of these objects. This will help uncover possible trends and enhance our understanding of the role of $^{12}$CO/$^{13}$CO as a formation tracer.

\begin{acknowledgements}
Support for this work was provided by NL-NWO Spinoza SPI.2022.004. Our work is based on observations collected at the European Organisation for Astronomical Research in the Southern Hemisphere under ESO programme 1110.C-4264. D.G.P., S.d.R. and I.S. acknowledge support from NWO grant OCENW.M.21.010. This research has made use of NASA’s Astrophysics Data System and the python packages NumPy \citep{Numpy2020}, SciPy \citep{Scipy2020}, Matplotlib \citep{Hunter2007}, petitRADTRANS \citep{Molliere2019}, PyAstronomy \citep{Czesla2019}, Astropy \citep{Astropy2022}, corner \citep{Foreman2016}.
\end{acknowledgements}

\bibliographystyle{aa}
\bibliography{aa54195-25.bib}

\begin{appendix}\onecolumn

\section{Validation retrieval results} \label{app:test}
\begin{figure*}[ht!]
    \centering
    \includegraphics[width=\linewidth]{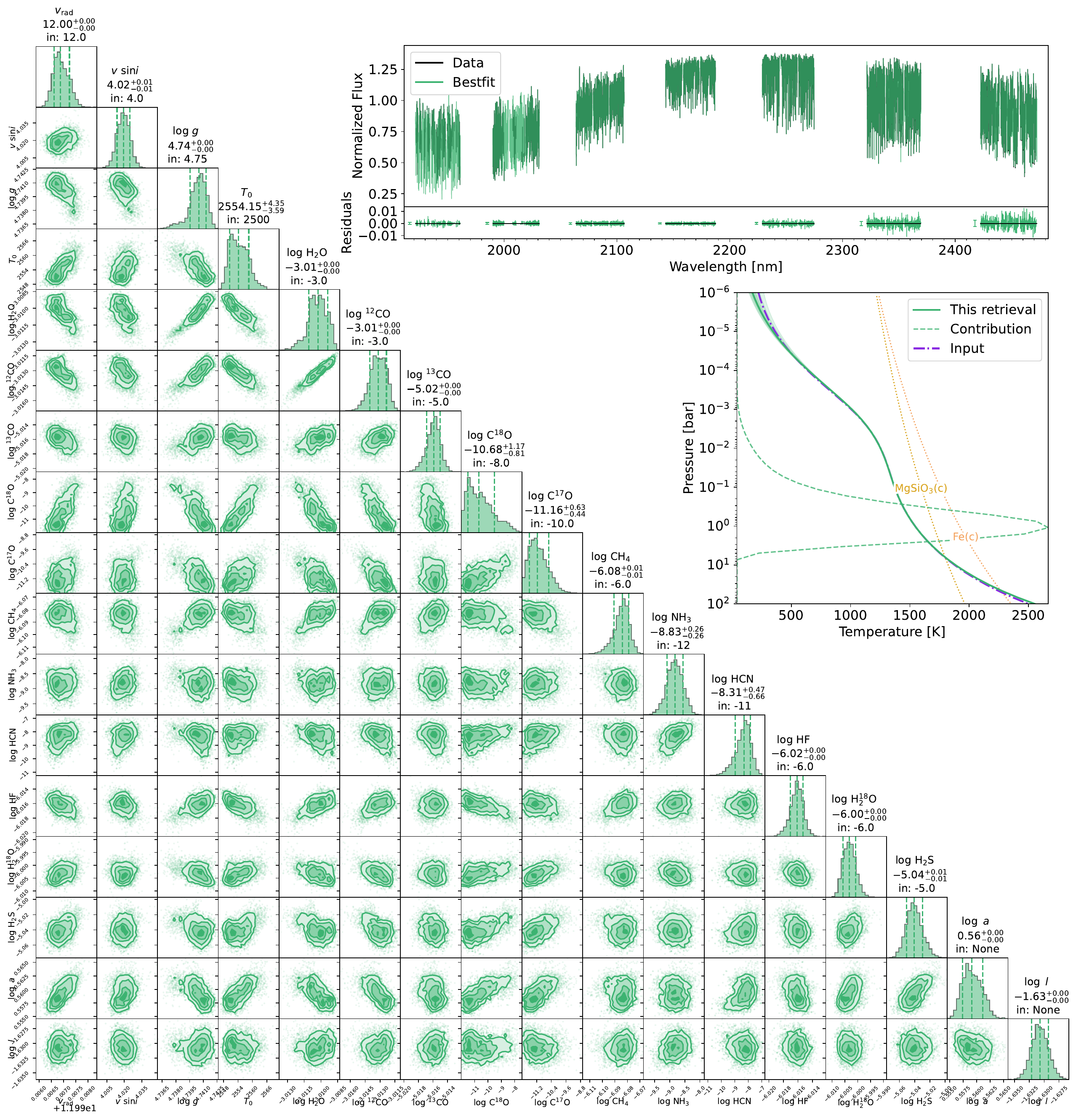}
    \caption{Summary of the validation retrieval on a synthetic test spectrum. Top right: Best-fit spectrum (green) compared to the input spectrum (black), including their residuals (bottom panel). Middle right: Retrieved $P$-$T$ profile (green) compared to the input $P$-$T$ profile (purple). Bottom left: Retrieved posterior distributions of the free parameters. Parameter names in the titles on the diagonal, with the retrieved values underneath, and the input value (signified with "in:") underneath.}
\end{figure*}
\clearpage

\section{Best-fitting spectra} \label{app:bestfit}
\begin{figure*}[ht!]
    \centering
    \includegraphics[width=0.9\linewidth]{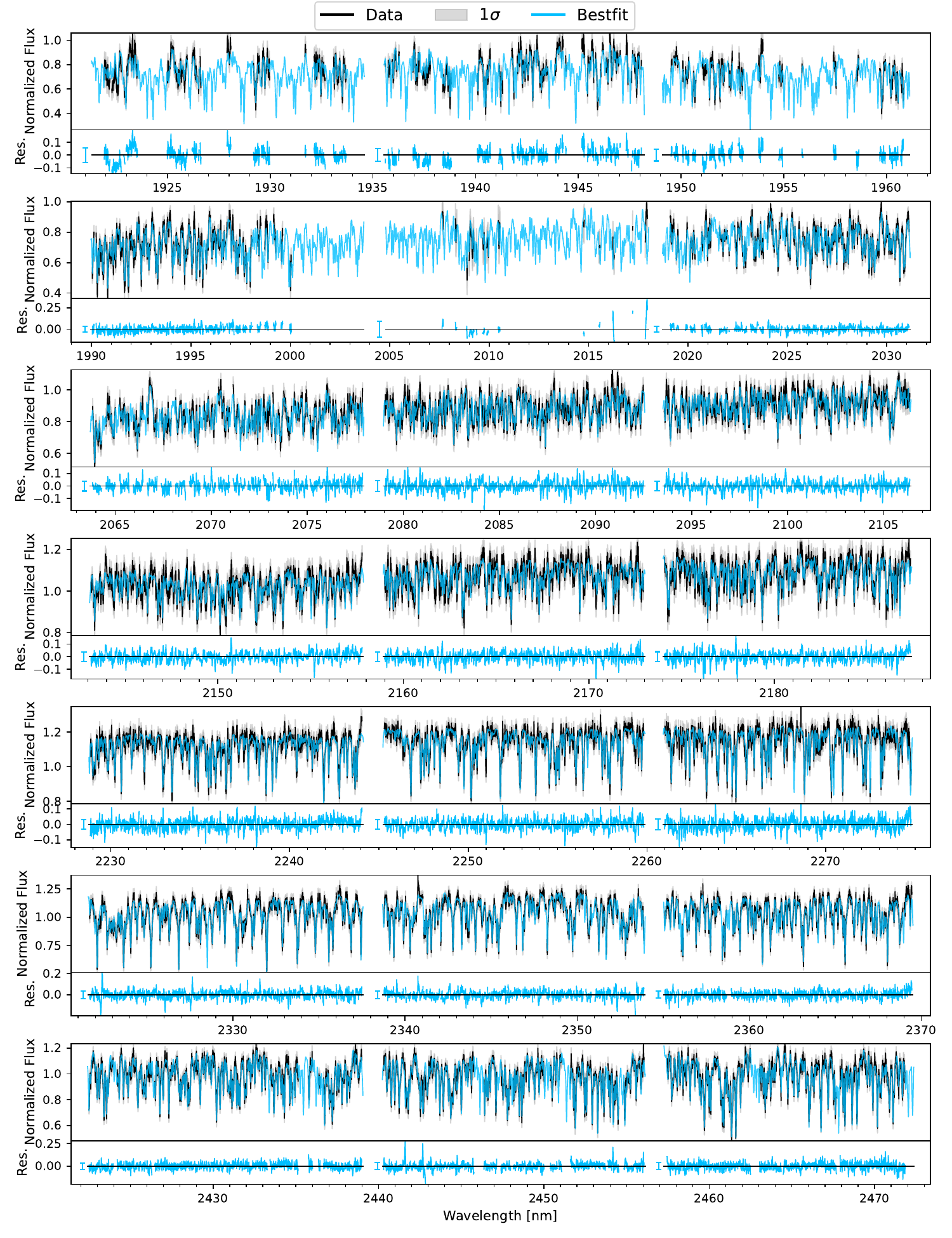}
    \caption{Best-fit model for 2M0355, with each panel showing a different spectral order. The observed spectrum is shown in black, along with its 1\,$\sigma$ uncertainties in gray, and the best-fit model in blue, with the residuals in the panel underneath each order.}
\end{figure*}

\begin{figure*}[ht!]
    \centering
    \includegraphics[width=0.9\linewidth]{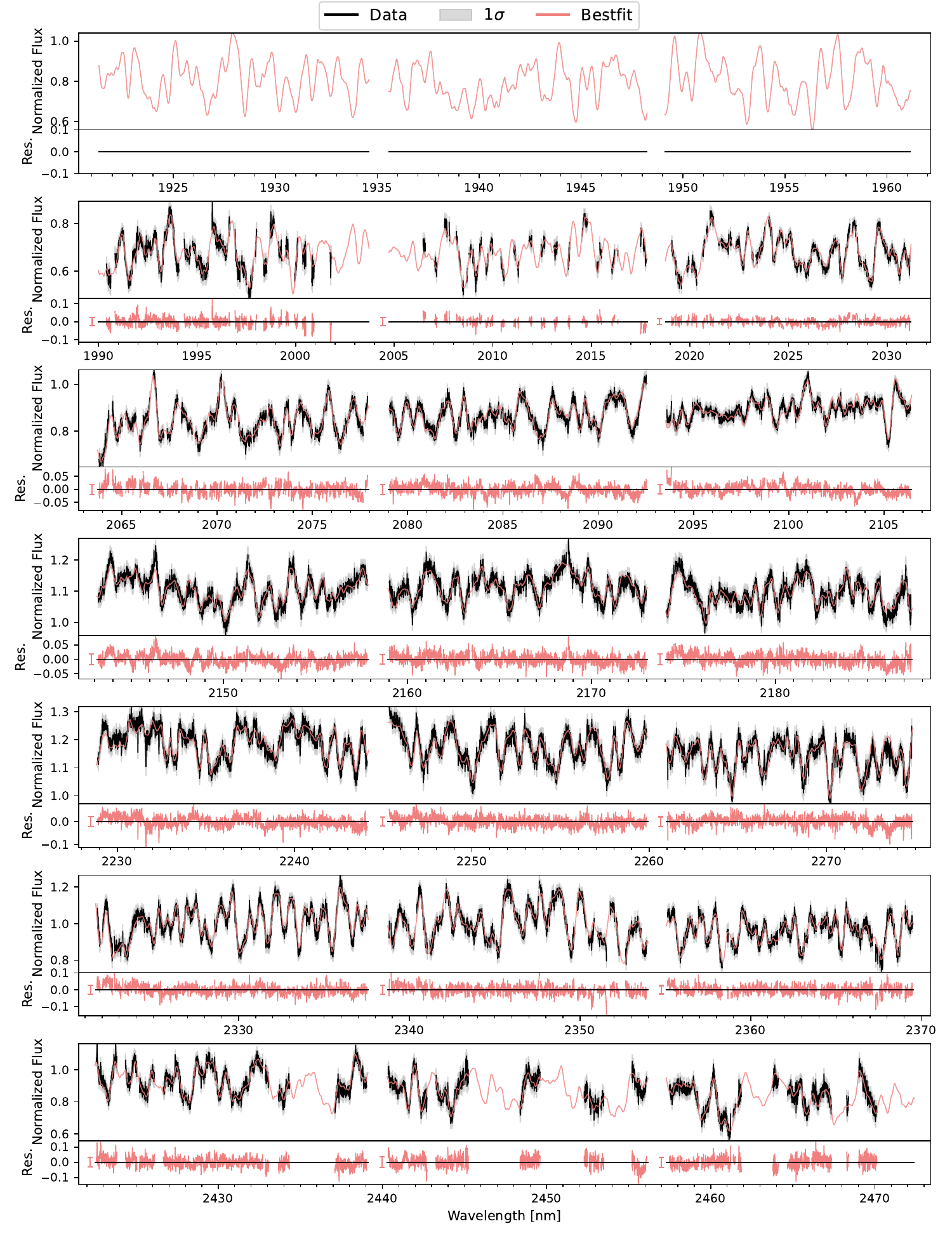}
    \caption{Best-fit model for 2M1425, with each panel showing a different spectral order. The observed spectrum is shown in black, along with its 1\,$\sigma$ uncertainties in gray, and the best-fit model in blue, with the residuals in the panel underneath each order.}
\end{figure*}

\clearpage

\section{Extended cornerplot} \label{app:cornerplot_rest}
\begin{figure*}[ht!]
    \centering
    \includegraphics[width=\linewidth]{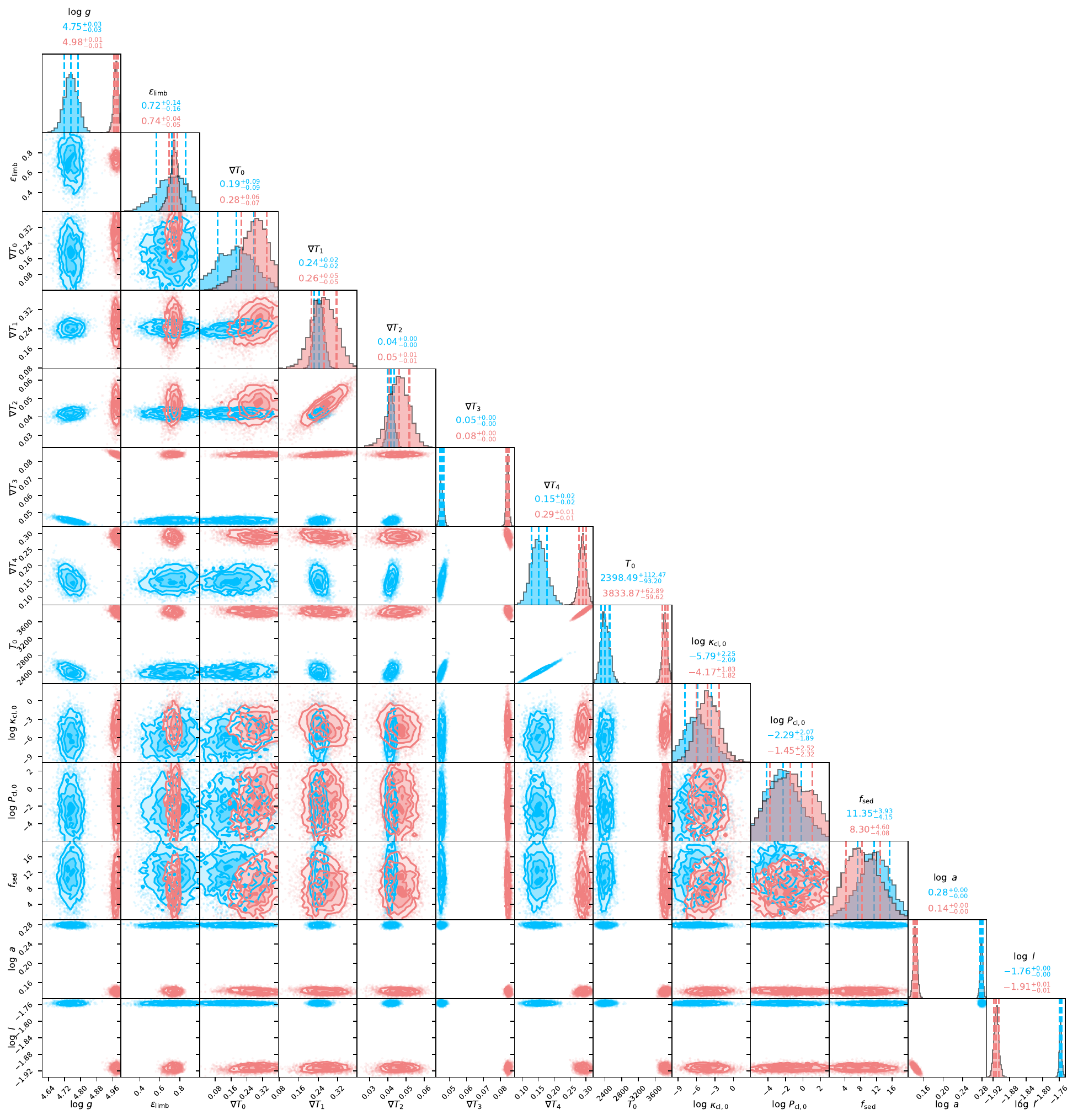}
    \caption{Posterior distributions of the remaining retrieved free parameters not shown in Figure~\ref{fig:summary} for 2M0355 (blue) and 2M1425 (pink). Due to the large differences in $v_{\text{rad}}$ and $v\,\text{sin}\,i$, they are omitted from the plot.}
\end{figure*}

\clearpage

\section{Comparison of free and equilibrium chemistry} \label{app:free_equ}

\begin{figure*}[h!]
    \centering
    \begin{subfigure}{0.5\textwidth}
        \centering
        \includegraphics[width=\linewidth]{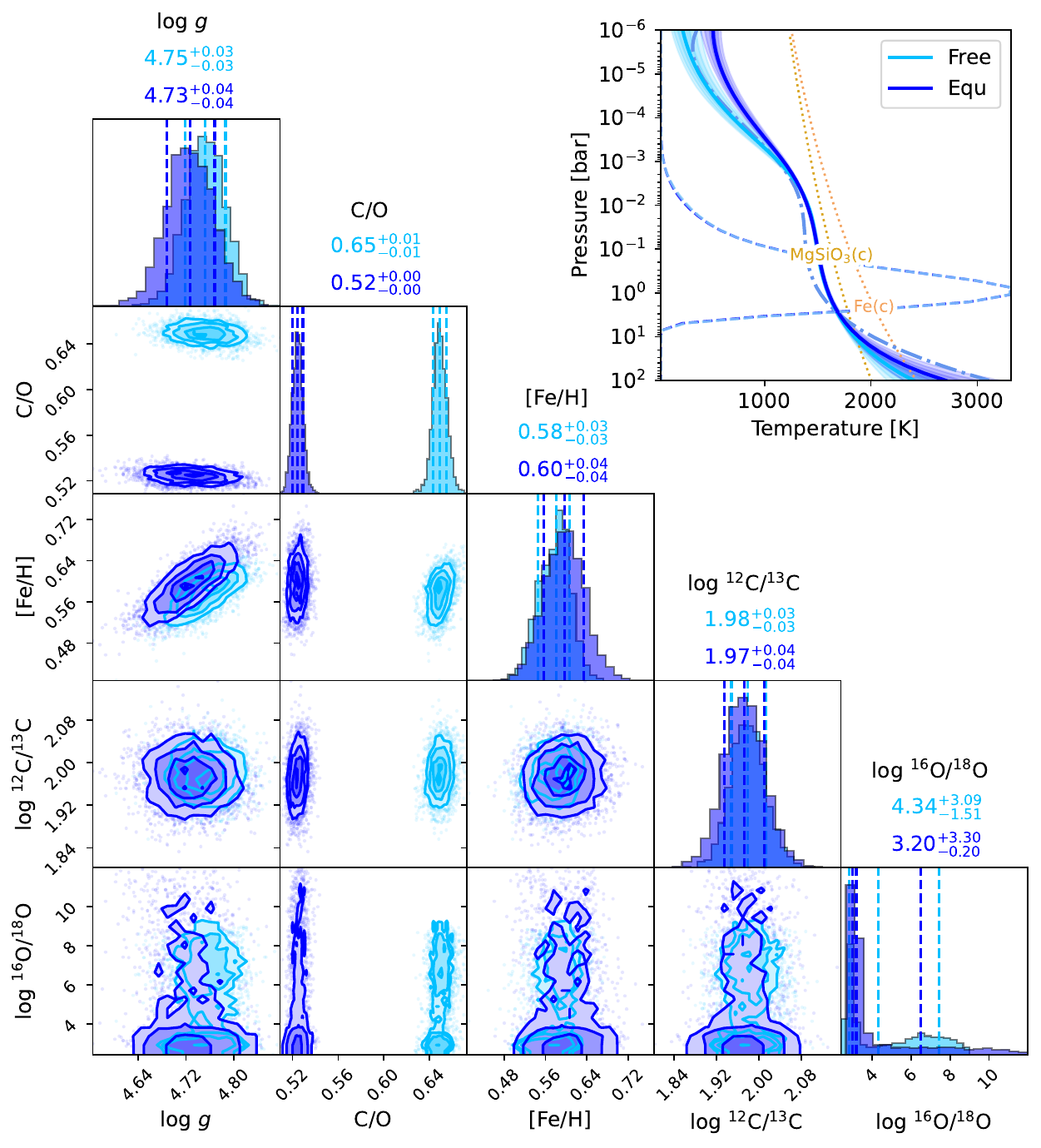}
        \caption{2M0355}
    \end{subfigure}
        \vspace{0.5cm}
    \begin{subfigure}{0.5\textwidth}
        \centering
        \includegraphics[width=\linewidth]{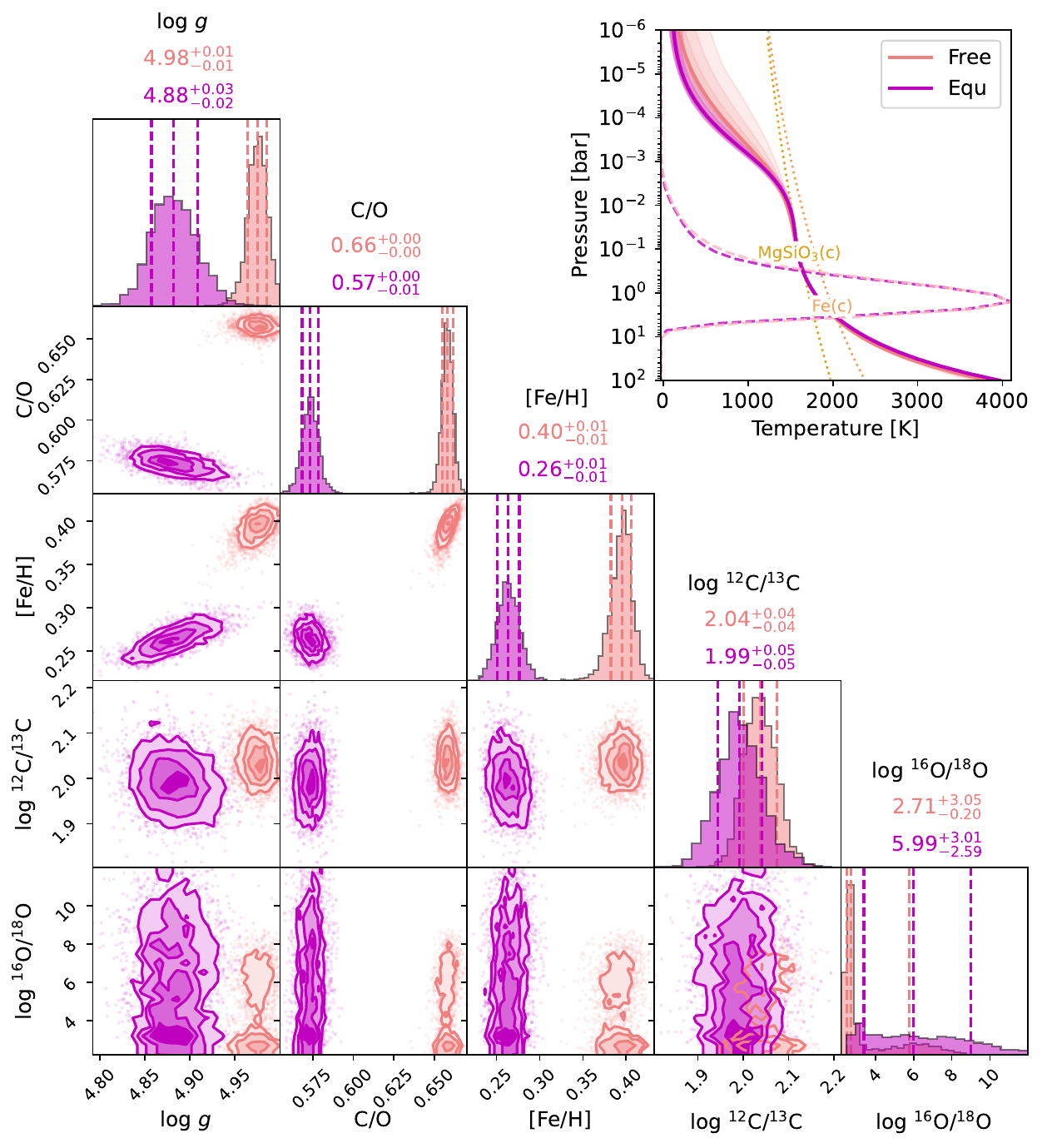}
        \caption{2M1425}
    \end{subfigure}
    \caption{Comparison of free and equilibrium chemistry retrieval results for 2M0355 (top) and 2M1425 (bottom). Bottom left: posterior distributions of selected parameters. Top right: $P$--$T$ profiles and their corresponding emission contribution functions.}
    \label{fig:free_equ_corner}
\end{figure*}

\end{appendix}
\end{document}